\documentclass[12pt]{article}
\pdfoutput=1
\usepackage{amsmath,amssymb,graphicx} %use drftcite in draftmode
\usepackage{epsf}
\usepackage{pstricks}
\usepackage[sort&compress,numbers]{natbib}
\usepackage{hyperref}
\usepackage{ctable}
\usepackage{multirow}

\newcommand{\beq}{\begin{eqnarray}}% can be used as {equation} or  {eqnarray}
\newcommand{\eeq}{\end{eqnarray}}

\newcommand\ZZ{\hbox{\zfont Z\kern-.4emZ}}
\font\zfont = cmss10 %scaled \magstep1

\newcommand{\met}{\mbox{${\rm \not\! E}_{\rm T}$}}
\newcommand{\dzero}{D$\varnothing$~}
\newcommand{\dz}{D$\varnothing$}
\def\gappeq{\mathrel{ \rlap{\raise.5ex\hbox{$>$}}
                      {\lower.5ex\hbox{$\sim$}}  } }
\def\lappeq{\mathrel{ \rlap{\raise.5ex\hbox{$<$}}
                      {\lower.5ex\hbox{$\sim$}}  } }
\begin{document}
\begin{titlepage}
\begin{flushright}
{\small YITP-SB-10-25}
\end{flushright}

\vskip.5cm
\begin{center}
{\huge \bf Long-Lived Neutralino NLSPs
 \vspace{.2cm}}

\vskip.1cm
\end{center}
\vskip0.2cm

\begin{center}
{\bf Patrick Meade$^a$, Matthew Reece$^b$, and David Shih$^{c,d}$}
\end{center}
\vskip 8pt

\begin{center}
{\it $^a$C. N. Yang Institute for Theoretical Physics,\\ Stony Brook University, 
Stony Brook, NY 11794} \\
\vspace*{0.1cm}
{\it $^b$Princeton Center for Theoretical Science\\
Princeton University, Princeton, NJ 08544, USA}\\
\vspace*{0.1cm}
{\it $^c$Institute for Advanced Study\\
Princeton, NJ 08540} \\
\vspace*{0.1cm}
{\it $^d$ Department of Physics and Astronomy, \\
Rutgers, 136 Frelinghuysen Rd, Piscataway, NJ 08854}\\
\vspace*{0.1cm}

\end{center}

\vglue 0.3truecm

\begin{abstract}
\vskip 5pt \noindent  

We investigate the collider signatures of heavy, long-lived, neutral particles that decay to charged particles plus missing energy. Specifically, we focus on the case of a neutralino NLSP decaying to $Z$ and gravitino within the context of General Gauge Mediation. We show that a combination of searches using the inner detector and the muon spectrometer yields a wide range of potential early LHC discoveries for NLSP lifetimes ranging from $10^{-1}-10^5$ mm. We further show that events from $Z(\ell^+\ell^-)$ can be used for detailed kinematic reconstruction, leading to accurate determinations of the neutralino mass and lifetime. In particular, we examine the prospects for detailed event study at ATLAS using the ECAL (making use of its timing and pointing capabilities) together with the TRT, or using the muon spectrometer alone. Finally,
we also demonstrate that there is a region in parameter space where the Tevatron could potentially discover new physics in the delayed $Z(\ell^+\ell^-)+\met$ channel. While our discussion centers on gauge mediation, many of the results apply to any scenario with a long-lived neutral particle decaying to charged particles. 

\end{abstract}

\end{titlepage}

\newpage

\renewcommand{\thefootnote}{(\arabic{footnote})}

%%%%%%%%%%%%%%%%%%%%%%%%%%%%%%%%%%%%%%%%%%%%%%%%%%%
\section{Introduction}\label{sec:intro}
\setcounter{equation}{0} \setcounter{footnote}{0}
%%%%%%%%%%%%%%%%%%%%%%%%%%%%%%%%%%%%%%%%%%%%%%%%%%%

In this work, we will study the collider signatures of neutral, heavy, long-lived particles that decay to multiple charged particles and missing energy. Such signatures are interesting for a variety of reasons. First, the combination of displaced decays, significant missing energy, and high invariant mass generally leads to extremely clean, nearly background-free channels. So such signatures can be excellent discovery channels for new physics. Second, the presence of multiple charged particles generally allows for detailed reconstruction of the event kinematics, including the position and time of the displaced vertex. This in turn can be used to deduce the properties of the neutral parent particle, including its mass and lifetime.

The prototypical scenario we have in mind is gauge-mediated SUSY breaking (GMSB) with a long-lived general neutralino NLSP.\footnote{Although our discussion focuses on GMSB for concreteness, many of our results will apply to more general scenarios with long-lived neutral particles decaying to charged particles, such as Hidden Valleys \citep{hv1,hv2}.}
 The general neutralino NLSP is a mixture of bino, Higgsino, and wino gauge eigenstates. If the NLSP has a significant Higgsino or wino component, then it can have ${\mathcal O}(1)$ branching fractions to $Z+\tilde G$ and $h+\tilde G$, where $\tilde G$ is the gravitino. Subsequent decays of the $Z$ and $h$ can then give rise to the signatures involving charged particles and missing energy. In particular, $Z\to e^+e^-$, $Z\to \mu^+\mu^-$, $Z\to \tau^+\tau^-$, $Z\to {\rm jets}$ and $h\to b\bar b$ plus missing energy are all possible signatures. In this paper, we will focus primarily on the simplest and cleanest cases, namely $Z\to \ell^+\ell^-$ with $\ell=e$ or $\mu$. For the purposes of discovery with early LHC data, we will also consider $Z\to {\rm jets}$.

While Higgsino and wino NLSPs do not occur in minimal gauge mediation, they are perfectly reasonable possibilities in the context of General Gauge Mediation \cite{GGM}. Higgsino NLSPs also arise in the simplest extensions of minimal gauge mediation \cite{Cheung:2007es} and can alleviate fine-tuning \cite{Agashe:1997kn}. For a review of minimal gauge mediation and its collider phenomenology, and many original references, see e.g.\ \cite{GMSBReview}.
  
The NLSP lifetime is determined by the fundamental scale of SUSY breaking, $\sqrt{F}$. For $\sqrt{F}\lesssim 10^2$ TeV, the NLSP generally decays promptly, while for $\sqrt{F}$ between $\sim 10^2-10^3$ TeV, it mostly decays away from the primary vertex but inside the detector.
Promptly-decaying neutralino NLSPs were considered some time ago at the Tevatron \cite{Ambrosanio:1996jn,Dimopoulos:1996vz,Dimopoulos:1996fj,Dimopoulos:1996yq,BMPZ,MatchevThomas, Baer:1999tx,SUSYworkinggroup} and the LHC \cite{BMTW}. More recently in \cite{OurPromptPaper}, we have considered current Tevatron limits on and future prospects for discovery of promptly-decaying general neutralino NLSPs. In addition, the possibility of finding Higgsino NLSPs decaying promptly to boosted Higgses using jet substructure was recently discussed in \cite{Kribs:2009yh}.

By contrast, displaced $Z$'s from NLSP decays are not as well-studied. The possibility of such signatures was first raised in \cite{Dimopoulos:1996vz,Dimopoulos:1996fj,Dimopoulos:1996yq,MatchevThomas, SUSYworkinggroup}. Motivated by minimal gauge mediation (where ${\rm Br}(\tilde{\chi}_1^0\to Z+\tilde G)\ll 1$), Ambrosanio and Blair \cite{Ambrosanio:1999iu} examined this signature at a hypothetical 500 GeV $e^+e^-$ linear collider, focusing on its uses for the post-discovery study of supersymmetry. As far as we know, our paper is the first to study this signature in detail at hadron colliders, and to examine its potential as a discovery channel for new physics.

On the experimental side, long-lived neutralino NLSPs have received relatively little attention, save for the case of a bino NLSP (again, motivated by minimal gauge mediation), which decays dominantly to photons.  
Such decays have been searched for at CDF \cite{cdfgammajetmettiming} using the EM timing system installed early in Run II \cite{cdfemtiming,TobackWagner}. Preliminary studies of non-pointing photons have been carried  out by the CMS \cite{CMSnonpointing,CMSGMSB} and ATLAS collaborations \cite{Kawagoe:2003jv, Prieur05}. Both CMS and ATLAS explored the possibility of using their electromagnetic calorimeter (ECAL) to determine that photons were non-pointing (i.e.\ originate from a direction other than that of the primary vertex), while ATLAS additionally explored the use of timing information in the ECAL and reconstructed tracks from photon conversions. 

By adapting the ATLAS work on non-pointing photons to the case of $Z\to e^+e^-$, we will see that a combination of the ECAL and tracking can lead to precise determination of the neutralino mass and lifetime.  Such displaced decays allow us to reconstruct the full kinematics of an event in ways that are simply impossible in events where all decays are prompt, or where the decay products are neutral. For comparison, the study \cite{Kawagoe:2003jv} had to work with a longer a decay chain $\tilde{\ell} \to \ell \tilde{\chi}^0_1 \to \ell \gamma \tilde{G}$, use photon conversions for extra directional information, and make use of known masses to fully reconstruct events. (For another recent discussion of the use of long-lived particles for full kinematic reconstruction, see \cite{ChangLuty}.) 

We also investigate $Z(\mu^+\mu^-)$ final states using the muon spectrometer. These play a complementary role to $Z(e^+e^-)$ final states; by analyzing both electron and muon final states, we can cover a much wider range of possible NLSP lifetimes.  The muon chambers are at the largest distance from the interaction point, and at both LHC experiments full 3D tracking information is recorded.  Using this information, plus potential timing information from the muon system, a full kinematic reconstruction of standalone muons can be done in principle, extending the detailed event study to much longer lifetimes than is possible with electrons alone.  

In reconstructing events, we are limited at short lifetimes by the timing resolution, which at best is $\mathcal{O}(100\,\mathrm{ps})$.\footnote{At long lifetimes we are limited only by the acceptance of the muon chambers. For much longer lifetimes, one might consider using detectors far from the LHC \cite{Meade:2009mu}.} We stress that this is the best any proposed study can do for shorter lifetimes without having to rely upon additional kinematic information about the masses.  
However, if one is interested in discovery of new physics and not full event reconstruction, there is additional reach at short lifetimes, since here it is not necessary to use timing information at all.  Given that the decays we investigate come from a $Z$ or $h$, the high invariant mass or large separation of any charged particles produced at a displaced vertex eliminate backgrounds such as conversions. We will show that including information from  the inner silicon layers of an LHC detector opens up discovery possibilities for delayed decays that have hitherto not been stressed for the LHC.

The outline of our paper is as follows. In section \ref{sec:lhcgen}, we will discuss the detector geometry of ATLAS, as well as the ability of different detector components to perform specialized measurements useful for long-lifetime searches, such as arrival times or the direction of a particle passing through a subdetector. In section \ref{sec:discovery}, we discuss the prospects for a discovery at the LHC in early data (1 fb$^{-1}$ taken at 7 TeV). For discovery, we are not interested in detailed reconstruction of events so much as simply establishing a signal that is distinct from Standard Model backgrounds. By making use of the pixel detector, the TRT, or the muon spectrometer, in the $Z\to e^+e^-$, $Z\to \mu^+\mu^-$ and $Z\to {\rm jets}$ final states, we show that a range of six orders of magnitude in lifetimes can be covered, from $10^{-1}-10^{5}$ mm. In section \ref{sec:atlasrecon}, we turn to the prospects for doing more detailed event reconstruction with the ATLAS detector using 10 fb$^{-1}$ of data at 14 TeV. Either the ECAL together with the TRT or the muon spectrometer alone allow full kinematic reconstruction of decays based on timing and directional information. We show that masses, lifetimes, and angular distributions can be measured accurately. In section \ref{sec:discussion}, we give a discussion of our results, including how they generalize to CMS. Appendix \ref{sec:tevatron} gives a summary of existing Tevatron searches for long lifetime and presents a possible search that could discover long-lived NLSPs at \dz.

\section{Long-Lived Neutral Particles at ATLAS}
\label{sec:lhcgen}

\begin{figure}[!t]
\begin{center}
\includegraphics[width=0.6\textwidth]{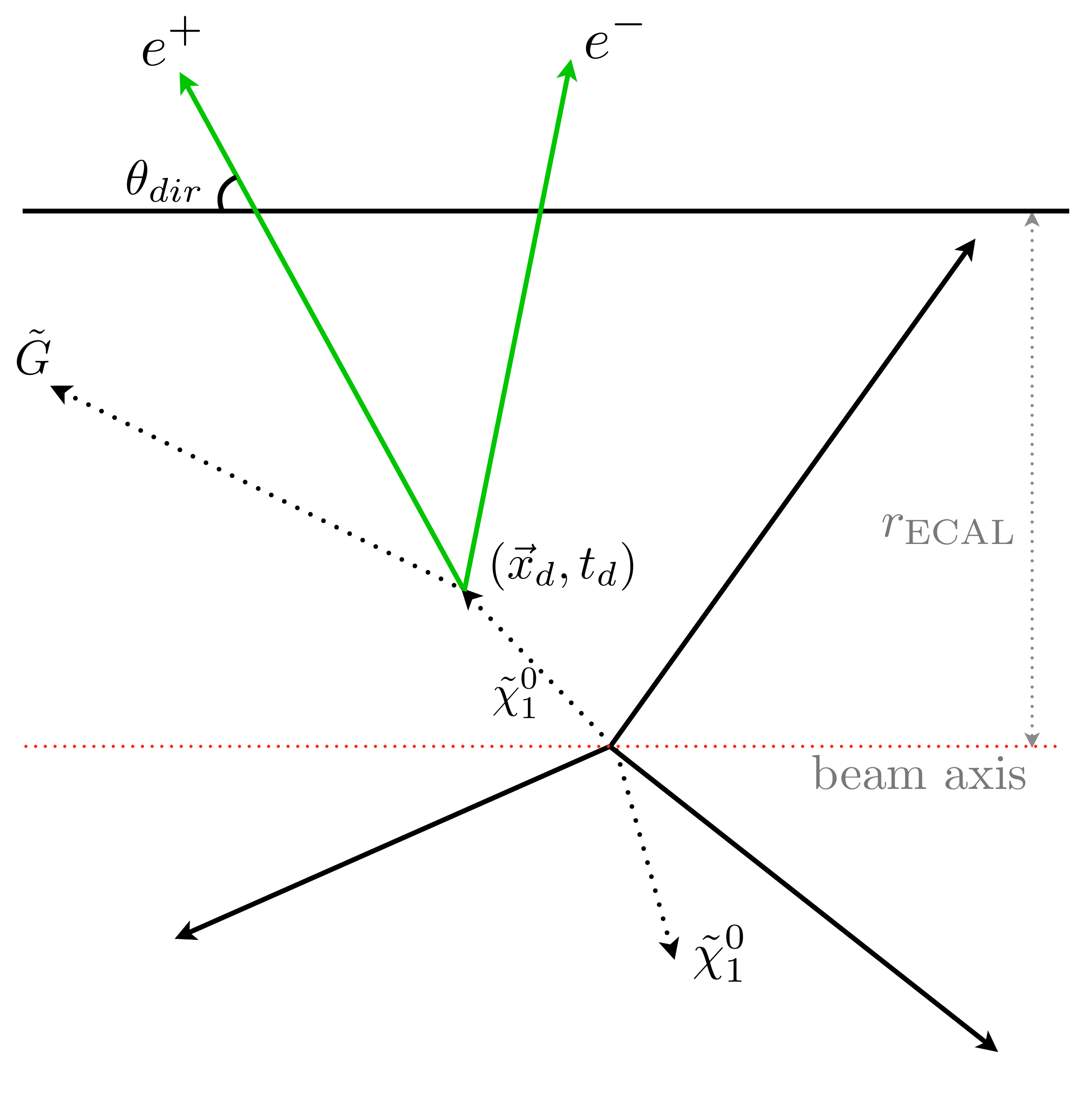}
\end{center}
\caption{A cartoon of what an interesting long-lifetime event could look like in a detector. The neutralino NLSP travels for a while and then decays to $Z(e^+e^-)+$gravitino. The solid black lines  denote additional jets, tracks, etc.\ that can be used to find the primary vertex. These can come from e.g.\ initial state radiation or decays of directly-produced colored sparticles to the NLSP. }
\label{fig:cartoon}
\end{figure}

In the rest of the paper, we will discuss the capabilities for ATLAS to identify displaced NLSP decays into $Z(e^+e^-)$ or $Z(\mu^+\mu^-)$ (and in section \ref{sec:discovery} also $Z\to {\rm jets}$) along with a gravitino.  These can serve as benchmark scenarios for more general long-lived neutral particle decays.  A schematic view of an interesting decay of this type is shown in Figure~\ref{fig:cartoon}.

%%%%%%%%%%%%%%%%%%%%%%%%%%%%%%%%%%%%%%%%%%%%%%%%%%%
\subsection{ATLAS Detector Geometry}
\label{sec:geom}
\setcounter{equation}{0} \setcounter{footnote}{0}
%%%%%%%%%%%%%%%%%%%%%%%%%%%%%%%%%%%%%%%%%%%%%%%%%%%

\begin{figure}[!t]
\begin{center}
\includegraphics[width=0.8\textwidth]{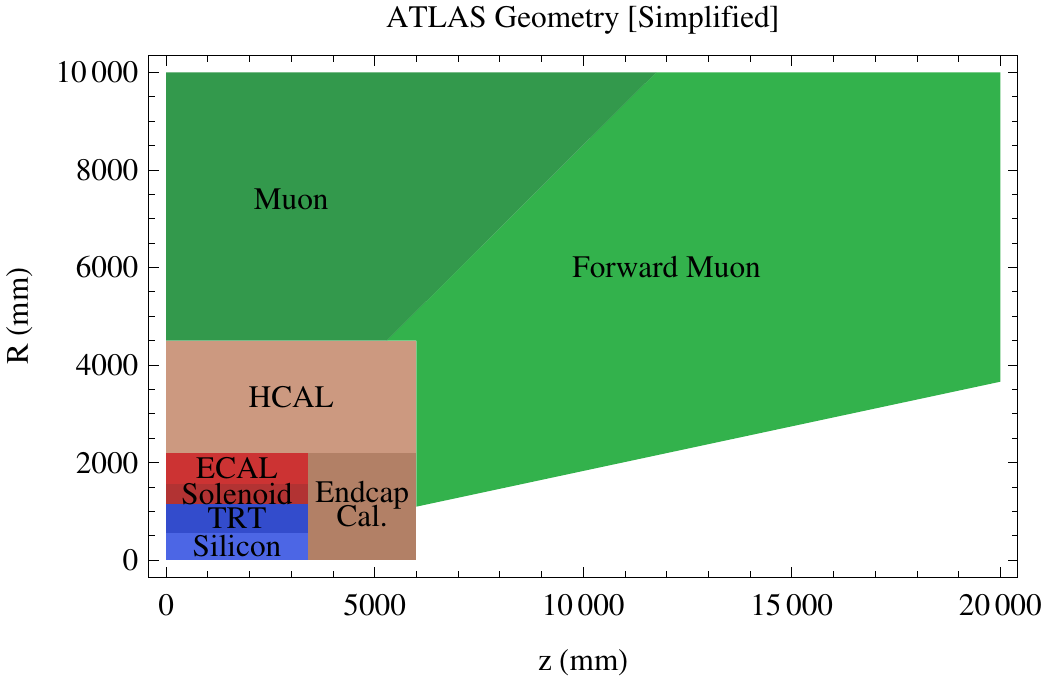}
\end{center}
\caption{Simplified illustration of one quadrant of the ATLAS detector geometry.}
\label{fig:ATLASgeom}
\end{figure}

In scenarios with a late-decaying particle, the collider signatures are determined by the interplay of kinematics and geometry. We show an example of such a geometry in Figure \ref{fig:ATLASgeom}. Note that the muon detectors occupy a large volume, while the inner detectors and tracking systems extend out to about a meter away from the beamline and a few meters along the beamline.

\begin{figure}[!t]
\begin{center}
\includegraphics[width=0.5\textwidth]{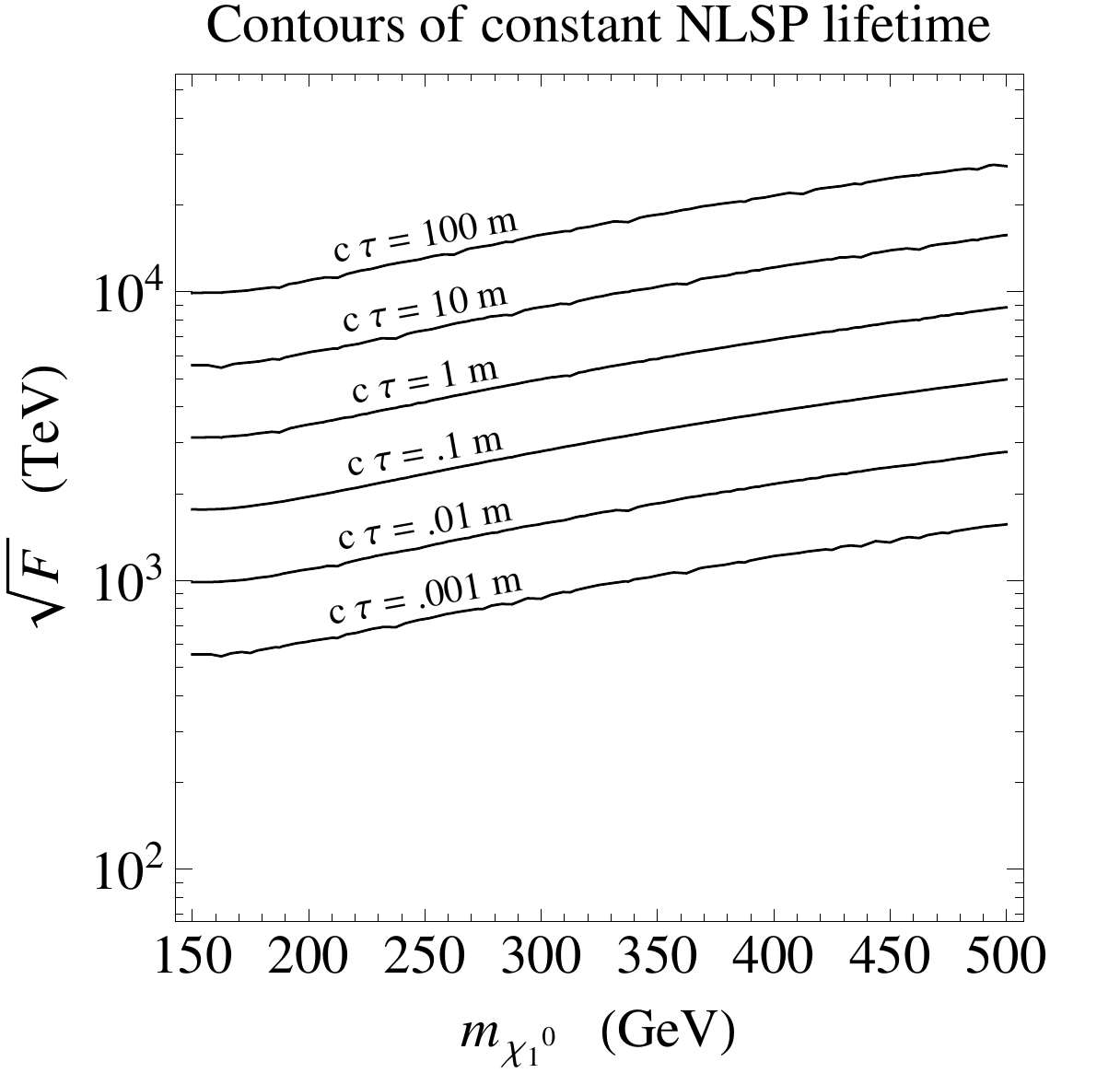}\includegraphics[width=0.5\textwidth]{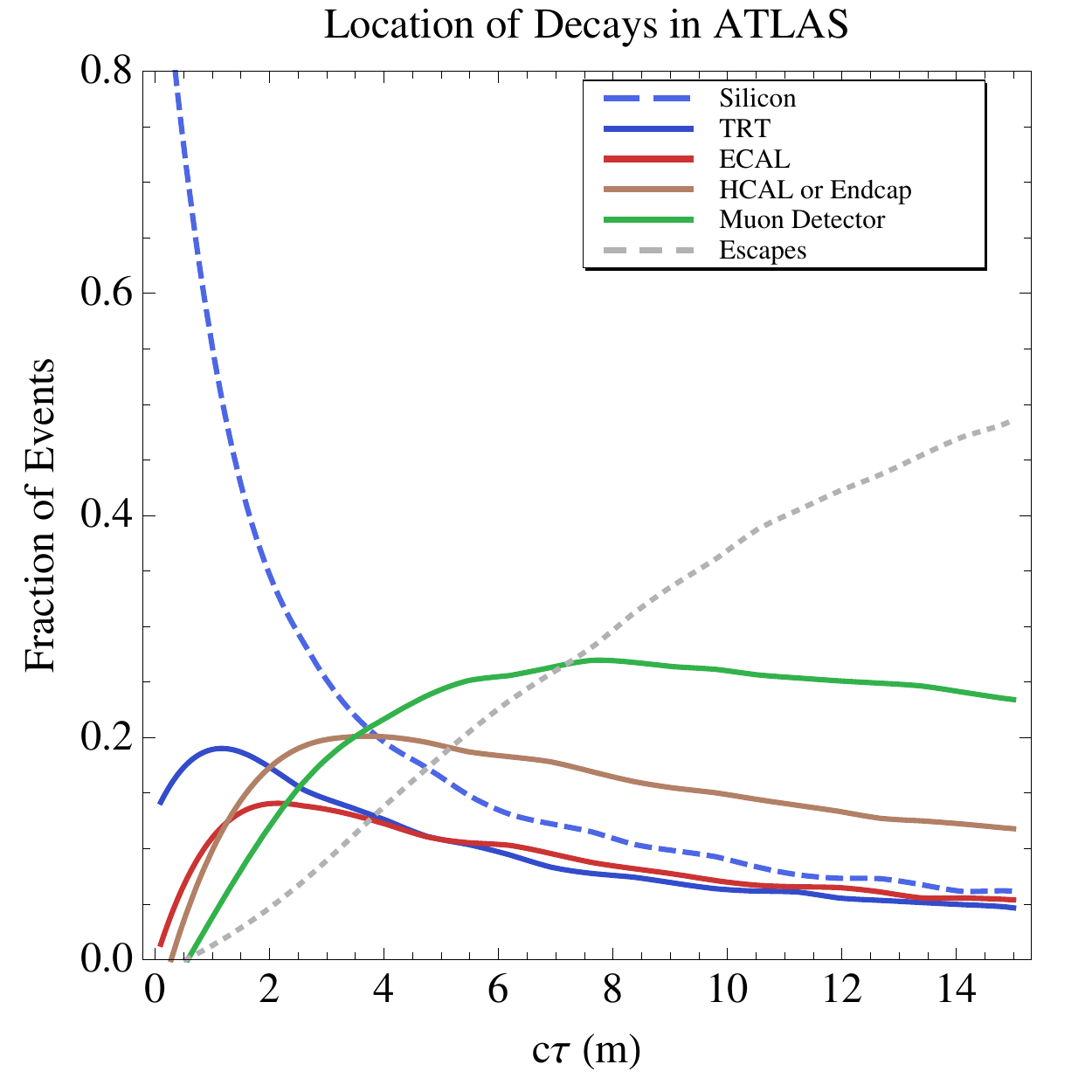}
\end{center}
\caption{Left plot: contour plot of Higgsino NLSP proper lifetime vs.\ NLSP mass and SUSY-breaking scale. Right plot: Fraction of decays located in different subdetectors at ATLAS, as a function of $\tilde{\chi}^0_1$ lifetime. Here the NLSP was taken to have  $m_{NLSP}= 250$ GeV and was assumed to come from decays of 600 GeV gluinos. However, the dependences on NLSP and gluino mass are not very strong compared to the effects of $c\tau$ and detector geometry.}
\label{fig:decayloc}
\end{figure}

In gauge mediation, the overall scale of the NLSP decay width is set by a dimensionful quantity
${\cal A}$:
\begin{equation}
\label{eqn:decaylength}
{\cal A} = \frac{m^5_{\tilde{\chi}^0_1}}{16\pi F^2} \approx \left(\frac{m_{\tilde{\chi}^0_1}}{100~{\rm GeV}}\right)^5 \left(\frac{100~\rm{TeV}}{\sqrt{F}}\right)^4 \frac{1}{0.1~{\rm mm}}.
\end{equation}
Here $\sqrt{F}$ is the fundamental scale of SUSY breaking; it is related to the gravitino mass via $m_{3/2} = F/(\sqrt{3}M_{Pl})$. In Figure \ref{fig:decayloc}, we plot (at left) the relationship between lifetime and SUSY-breaking scale and (at right) the fraction of decays occurring within different regions of the detectors at ATLAS as a function of the lifetime. (A plot similar to the right-hand plot, generated for hidden-valley models, has appeared in refs. \cite{DanVentura,TriggeringATLAS}.) The general pattern is this: at short lifetimes, nearly all of the decays occur within the inner detector. Consequently, nearly all the decays leave tracks in the TRT. As the lifetime becomes of order meters, 10-20\% of the decays can occur in the outer layers of the detector: the electromagnetic or hadronic central calorimeters, the forward calorimeters, or the muon detectors. However, at very long lifetimes,  most NLSPs will escape the detector altogether. The distribution of decay locations for those NLSPs that decay inside the detector is essentially flat, i.e.\ it depends only on the relative volumes of the different subdetectors. 

Analogous plots can be made for CMS and for the Tevatron detectors. Essentially the only qualitative difference these have with fig.\ \ref{fig:decayloc} is the fraction of decays in the muon detectors. At ATLAS, the very large volume of the muon system allows it to surpass the number of decays in the inner detector for lifetimes of around 6 meters or more. At the other detectors, decays in the muon system are somewhat less common, but still a useful tool at long lifetime.

The general lesson we learn here is that if a substantial fraction of events have NLSP decays anywhere in the detector, they will have NLSPs decaying in the inner detector. Thus we expect that the decay products will pass through the calorimeters, and potentially leave tracks. It is also interesting to look for anomalous tracks in the muon detectors. Decays in the ECAL or HCAL are comparatively rare, and they offer less precision directional information, so we will not pursue the use of such decays for reconstructing events. Of course, ultimately, one would hope to find decays in all parts of the detector, and use their relative rates to help characterize lifetimes.

\subsection{ATLAS Detector Capabilities}
\label{sec:detcap}

\subsubsection{$Z\to e^+e^-$: TRT and ECAL}

When focusing on $Z(e^+e^-)$, the goal is to use as much precision information as possible from the ECAL and the TRT to reconstruct the decay chain and fit masses and other kinematic information. There are two sets of resolutions we will be interested in: those pertaining to how well the ECAL can measure energy, timing, and pointing information, and how well the TRT can in principle be used for finding displaced vertices.  We will use resolutions reported in the ATLAS note \cite{Prieur05} (see also \cite{ATLAStime}), which is a detailed follow-up on an earlier paper examining the capability of ATLAS to find non-pointing photons in minimal gauge mediation \cite{Kawagoe:2003jv}. 

\begin{table}[t!]
\begin{center}
\
\begin{tabular}{|c|c|c|} \hline
 & Measurement & Resolution\\\hline
\multirow{4}{*}{ECAL} &  $E$ &  $\delta E \sim 0.1 \sqrt{E~{\rm GeV}}$ \\
 & $\eta_{det}, \varphi_{det}$ & $\sigma_\eta = 0.004/\sqrt{E/{\rm GeV}}, \sigma_\varphi = 0.005/\sqrt{E/{\rm GeV}}$\\
 & $\theta_{dir}$ & $\sigma_\theta = \left(0.080 + \frac{|z_{e.v.}|}{100~{\rm cm}} 0.340 \right) / \sqrt{E/{\rm GeV}}$\\
 & $t_{det}$ &  $\sigma_t = 100$ ps \\
\hline
\multirow{1}{*}{TRT} & $\varphi_{dir}$ &  $\sigma_{\varphi_{dir}} = 1~{\rm mrad}$\\
\hline
\multirow{3}{*}{Muon} & $p$ & $\sigma_p= 0.04 p$ \\
 & $\theta_{dir}$,$\varphi_{dir}$ & $\sigma_{\varphi_{dir}} = \sigma_{\theta_{dir}} = 15~{\rm mrad}$\\
 & $t_{det}$ & $\sigma_t = 2$ ns \\ \hline
\end{tabular}
\end{center}
\caption{Measured parameters and their resolutions in the ATLAS detector. The $det$ subscripts refer to a position or absolute time measured in the detector.  The $dir$ subscripts refer to the direction of the energy/momentum as measured by the detector. The ``effective $z$-vertex" $z_{e.v.}$ is found by extrapolating the particle's direction back in the $z - r$ plane to the point at which it intersects the $r = 0$ axis.}
\label{table:resolutions}
\end{table}

With the ECAL alone, we measure the five quantities listed in Table \ref{table:resolutions}. Given $Z \to e^+ e^-$, we thus measure ten numbers characterizing the two electrons. Up to discrete ambiguities and measurement errors, one can think of these numbers as equivalent to two three-momenta ${\bf p}(e^+)$ and ${\bf p}(e^-)$, together with the four coordinates of the decay vertex, $(x_d, y_d, z_d, t_d)$. In particular, the two timing measurements and two pointing angle measurements can be used to solve for the four decay coordinates $(x_d, y_d, z_d, t_d)$; ${\bf p}(e^+)$ and ${\bf p}(e^-)$ then point from the decay vertex to the calorimeters and have magnitude determined by the energy measurement. With the further assumption that we have a decay $\tilde{\chi}^0_1 \to Z \tilde{G} \to e^+ e^- \tilde{G}$, where $\tilde{G}$ is massless, this allows us to fully reconstruct the kinematics of the NLSP decay.\footnote{Dropping the assumption of a massless gravitino, the quantity that is measured is the $Z$ boson energy in the neutralino rest frame, $E^{rest}_Z = \frac{1}{2 m_{\tilde{\chi}^0_1}} \left(m^2_{\tilde{\chi}^0_1} + m^2_Z - m^2_{\tilde{G}}\right).$ This shows that in general, a degeneracy remains, so that $m_{\tilde{\chi}^0_1}$ and $m_{\tilde G}$ can both be increased consistent with all measurements.}

We are not limited to the ECAL for measurements of $Z(e^+e^-)$. We assume that ATLAS will be able to measure some tracking information for the $e^+$ and $e^-$ using the barrel TRT.  The barrel TRT is only able to measure the direction of charged particles in the transverse ($r - \varphi$) plane with a resolution listed in Table~\ref{table:resolutions}.  The use of the TRT to gain more directional information will possibly require additional experimental work.  Tracks can be constructed from an ``outside-in" tracking algorithm, although the description in Ref. \cite{ATLAS} suggests that tracks are only reconstructed by this algorithm if they point back to the beamline. Nonetheless, we expect that if there are sufficiently many TRT hits, a simple modification of the Hough transform algorithm \cite{DudaHart} can find the tracks of our electrons, especially with the constraints that the tracks are nearly straight and hit the ECAL at the location of the energy deposits.

\subsubsection{$Z\to \mu^+\mu^-$: Muon spectrometer}

For the case of $Z(\mu^+\mu^-)$, we are interested in the resolution achieved for standalone muon reconstruction (without requiring a matched TRT track). We have estimated these resolutions on the basis of information in the ATLAS muon TDR \cite{ATLASmuonTDR}, Table 6.1 of reference \cite{Aad:2008zzm} and discussions with ATLAS experimentalists \cite{JohnHobbs}. In Table~\ref{table:resolutions} we collect the various resolutions used in the study for $Z(e^+e^-)$ and $Z(\mu^+\mu^-)$.

Much like the combination of ECAL and TRT, the muon spectrometer provides 3d directional and timing information. We assume that its position measurements ($\eta_{det}$ and $\varphi_{det}$) have negligible error relative to directional measurements ($\theta_{dir}$ and $\varphi_{dir}$). Thus, the muon spectrometer alone provides a set of measurements that allow reconstruction comparable to that of the ECAL+TRT. Note that its time resolution is worse than that of the ECAL, but (being located at larger radius and encompassing a large volume) it is sensitive to decays with longer lifetime, so this is acceptable. Furthermore, unlike the TRT, the muon spectrometer's default algorithm should already be able to find 3D muon tracks without matching to hits in the inner detector.

\section{Discovery Potential with Early LHC Data}
\label{sec:discovery}

In this section, we will consider the capabilities of the ATLAS detector to discover new physics through displaced $Z\to e^+e^-$, $Z\to\mu^+\mu^-$ and $Z\to jj$ decays. We will focus on the discovery potential with early LHC data (1 fb$^{-1}$ at 7 TeV), primarily through direct production of colored sparticles. For this purpose, we have in mind a GMSB spectrum containing light-to-moderate mass squarks and/or gluinos which decay down to the neutralino NLSP (possibly through intermediate sparticles). The NLSP then decays with  Br$(\tilde{\chi}^0_1 \to Z \tilde{G}) \sim 1$. Such a spectrum does not occur in minimal gauge mediation, but as mentioned in the introduction, it is a perfectly viable possibility in General Gauge Mediation \cite{GGM}. 

\begin{table}[t!]
\begin{center}
\begin{tabular}{|c|c|}
\hline
\multicolumn{2}{|c|}{Cuts Shared by Discovery and Reconstruction Analyses:}\\
\hline
$|\eta_{det}| < 0.8$ & Passes through barrel TRT \\
$r_d < 800$ mm & Leaves sufficiently many TRT hits for track to be found\\
$\Delta R(e^+, e^-) > 0.4$ & Well-separated, unlike conversions\\
$E_T > 20$ GeV & Triggerable (2gamma20)\\
\hline
\multicolumn{2}{|c|}{Cuts Specific to Discovery with Silicon:} \\
\hline
$r_d < 50$ mm & Electrons pass through all Si layers \\
$r_d > 0.05$ mm & Reduce background \\
DCA $>0.05$ mm (either) & Reduce background \\
\hline
\multicolumn{2}{|c|}{Cuts Specific to Discovery with TRT:} \\
\hline
$r_d > 1$ cm & Reduce background \\
DCA $>1$ cm (either) & Reduce background \\
\hline
\multicolumn{2}{|c|}{Cuts Specific to Reconstruction with ECAL+TRT:}\\
\hline
$z_{e.v.} < 1200$ mm & Pointing resolution not too degraded \\
$\Delta t > 0.3$ ns & Significantly delayed \\
\hline
\end{tabular}
\end{center}
\caption{Cuts defining acceptance for $Z \to e^+ e^-$ analyses. Unless marked ``(either)", all cuts apply to {\em both} electrons in the decay.}
\label{table:accee}
\end{table}

\subsection{$Z\to e^+e^-$ with Si+TRT}

For early discovery with electrons, we consider an analysis analogous to the Tevatron searches described above: we use the tracker (TRT) to select events with significantly displaced decay radius $r_d=\sqrt{x_d^2+y_d^2}$. In addition, we use the fact that the silicon pixel detector at ATLAS is sufficiently large and precise that it can also be used to detect displaced $Z\to e^+e^-$ for shorter lifetimes. The combination of these can be used to cover lifetimes ranging from $10^{-1}$ to $10^{3}$ mm. 

The acceptance cuts used in this analysis are listed in detail in Table \ref{table:accee}. Some of them are shared with the more detailed reconstruction analysis we will present in the next section. They guarantee that the events leave tracks in the TRT, can be triggered on, and do not resemble conversion backgrounds.

There are additional cuts specific to using the TRT or Si layers to discover displaced decays, which impose a minimum radial decay distance $r_d$ as measured by the intersection of the electron tracks in the $r -\varphi$ plane. These cuts reduce backgrounds and select genuinely displaced decays. To be safe, we also require that at least one of the electron tracks (extrapolated backwards and forwards) satisfy a cut on the distance of closest approach (DCA) to the beamline. For the pixel analysis, we require $r>0.05$ mm and DCA$>0.05$ mm. This is based on the resolution of the pixel detector, which is given by the impact parameter resolution of about 0.01 mm reported in the ATLAS TDR \cite{ATLASTDRv1}. For the TRT analysis, we require $r$ and DCA$>1$ cm, again based on the TRT resolution in table \ref{table:resolutions}. We will refer to the fraction of events with a $Z \to e^+ e^-$ decay that pass these cuts as the ``acceptance."

Further background rejection could come from a $\met$ cut. These events will generically have large intrinsic \met, arising not only from the invisible gravitini and the escaping NLSP, but also because NLSP decay products hit the calorimeter at displaced locations relative to particles arising from the primary vertex. If there is significant strong production, the events will also contain hard jets from gluino or squark decays, giving another handle on backgrounds. While more detailed study from the experiment will be necessary to confirm that our signal can be isolated with high purity, we believe that (provided the experiment achieves its claimed resolutions) the signal would be dramatic and clean. 

\begin{table}[t!]
\begin{center}
\begin{tabular}{|c|c|}
\hline
\multicolumn{2}{|c|}{Cuts Shared by Discovery and Reconstruction Analyses:}\\
\hline
$r_d < 4500$ mm & Passes through all muon layers \\
$|\eta_{det}|<1.1$ at $r = 4.5,7.0,10.0$ m & Contained in the central muon spectrometer\\
Separation $>30$ mm at $r=4.5,7.0,10.0$ m & Resolve two muons\\
$r_d > 500$ mm & Significantly displaced vertex \\
$p_T > 20$ GeV & Triggerable\\
$\Delta t < 6$ ns (either) & Correct bunch-crossing ID \\
\hline
\end{tabular}
\end{center}
\caption{Cuts defining acceptance for $Z \to \mu^+ \mu^-$ analyses. Unless marked ``(either)", all cuts apply to {\em both} muons in the decay. We take $\Delta t$ to be measured at a radius of 7000 mm.}
\label{table:accmumu}
\end{table}%

\subsection{$Z\to \mu^+\mu^-$ with the muon system}

For $Z\to\mu^+\mu^-$ we focus on decays before the first layer of the muon system at $r=4500$ mm. The acceptance cuts are listed in Table \ref{table:accmumu}. Standalone muon tracks extrapolated back to the inner detector have an impact parameter resolution of $\sim 100$ mm, so we require a decay radius $r_d > 500$ mm, so that our muons are significantly displaced. (Furthermore, they will have no matched track in the silicon layers.) We drop the $\Delta R$ cut for muons, because conversions are less of a worry. However, a 30 mm minimum separation between the muons is imposed, and the muons cannot both arrive with too long a delay relative to promptly produced muons. If $\Delta t > 6$ ns for both muons, the muons are likely to be associated with the wrong bunch crossing \cite{DanVentura,TriggeringATLAS}. This imposes an upper bound to the range of lifetimes that can be probed with the muon spectrometer. Given all these cuts, we expect the background to be negligible.  

As we will see below, searching in this final state extends the lifetime reach roughly an order of magnitude beyond $Z\to e^+e^-$. Note that we only consider central barrel muons for this analysis, and the discovery potential could probably be further increased by using endcap muons. 

\subsection{$Z\to$jets}

We have focused on electrons and muons because they each give one clean track that can be measured well and used in reconstruction. However, for simply discovering new physics, one may wish to use decays to jets. Triggers that have been developed for Hidden Valley scenarios \cite{hv1,hv2} could prove useful for this purpose \cite{DanVentura,TriggeringATLAS,Giagu:2008im}.   We focus on a trigger path based on multiple ``regions of interest" in the muon spectrometer.   If a neutral object decays into jets after the HCAL, but before the first RPC layer in the muon system these events can be triggered on. We assume that if a $Z$ decays to jets in the muon spectrometer and is triggered on, there are enough handles in the event that backgrounds can be completely eliminated. Thus, to estimate the discovery potential in this channel, it suffices to estimate the trigger efficiency for the signal.

The trigger demands that there are at least 3 regions of interest in the muon system without a corresponding jet in the calorimeter, or tracks pointing to the same regions of interest.  Similar to the muon trigger, one of these regions of interest must have a $\Delta t<6$ ns so that the event is found in the correct bunch crossing.   Additionally a jet $E_T>35$ GeV is required.  As shown in~\cite{TriggeringATLAS} this trigger is approximately $70\%$ efficient for decays occurring between $4250$ and $7000$ mm in $r$. 

As we will show in the next subsection, a trigger-based analysis in this channel has the potential to extend the lifetime reach by another order of magnitude as compared to $Z\to\mu^+\mu^-$. This is because decays are allowed to happen out to 7000 mm for this trigger. Again we only utilize barrel-based muon regions of interest ($|\eta|<1$), but the acceptance can be further increased by using endcap muons ($|\eta|<2.5$).  Also, we have focused on a specific trigger pathway based on the muon system, but other triggers have been developed \cite{DanVentura,TriggeringATLAS,Giagu:2008im} which would be interesting to explore. For instance, one can use an additional jet based trigger where  $\log{E_{HAD}/E_{EM}}>1$ to find jets that originate from decays in the HCAL. This trigger has a similar efficiency to the region of interest trigger but for shorter lifetimes.

\subsection{Results}

\begin{figure}[!t]
\begin{center}
\includegraphics[width=.8\textwidth]{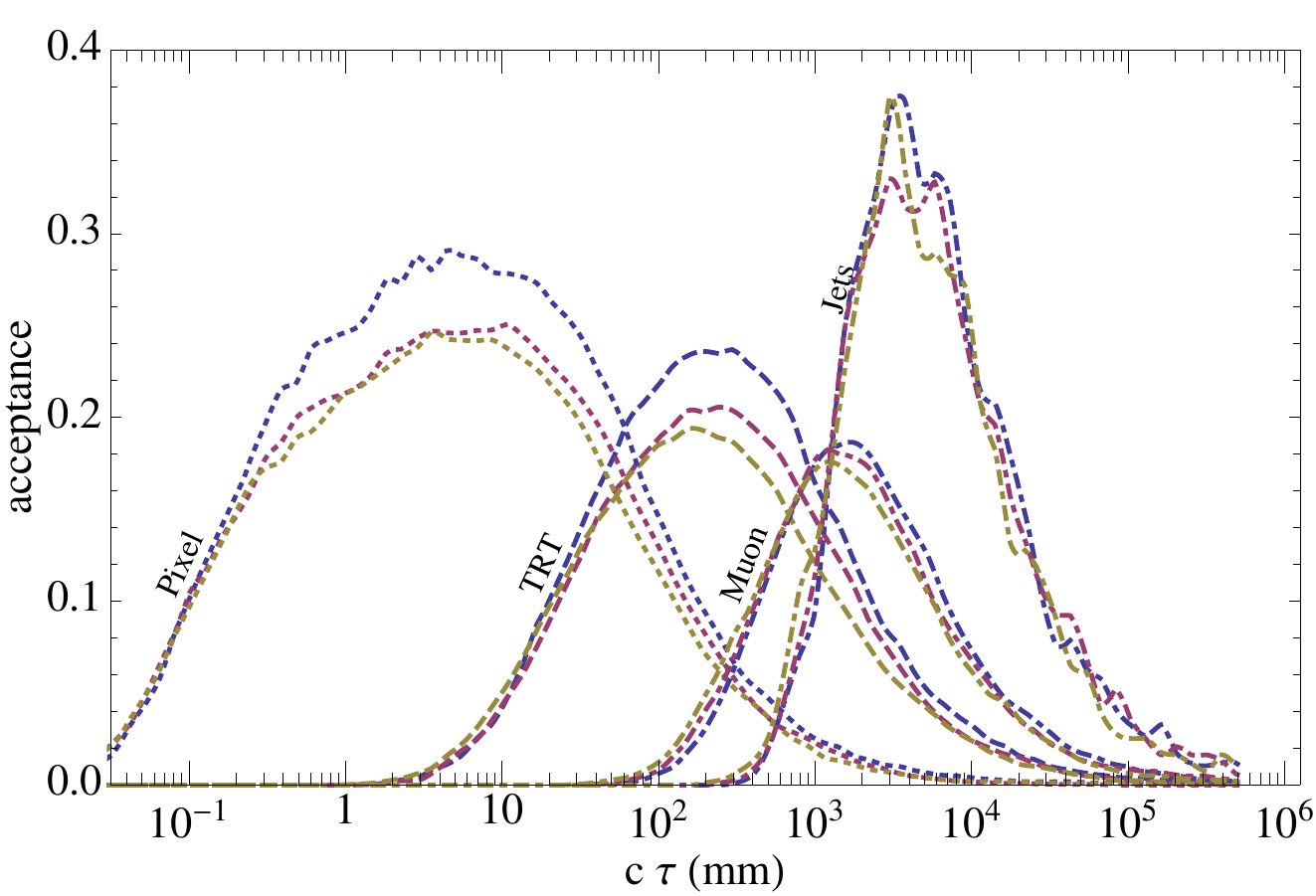}
\end{center}
\caption{Plots of the acceptances for the individual Si, TRT, muon, and jet based analyses as described in the text, as a function of the NLSP lifetime.   These acceptances have all been normalized to the number of $Z\to \mu^+\mu^-$ decays. The jet acceptance has been multiplied by the ratio of the branching fractions ${\rm Br}(Z\to jj)/{\rm Br}(Z\to\mu^+\mu^-)$.  The different curves (blue, red, yellow) correspond to different choices of the gluino mass $(M_{gluino}=600,\,800,\,1000)$.}
\label{fig:discacc}
\end{figure}

Now we put together the various analyses described above and estimate the discovery reach in early LHC running. Simulations in this subsection were performed with Pythia \cite{Pythia}.

In Figure \ref{fig:discacc}, we show how the acceptance changes as a function of lifetime, for different values of the gluino mass. We see that for a given analysis,  there is a slow loss of efficiency at longer lifetimes, as more NLSPs decay at too large a radius to give a signal in the relevant detector component (as expected from Figure \ref{fig:decayloc}). However, we also see that the analyses using the pixel detector, TRT and muon systems are nicely complementary to one another. Together, they provide coverage of lifetimes spanning $\sim 6$ orders of magnitude, from $\sim 10^{-1}$ to $10^{5}$ mm. This corresponds to slightly more than one order of magnitude in $\sqrt{F}$, from a few hundred to a few thousand TeV (cf. the left-hand plot of Figure \ref{fig:decayloc}).

\begin{figure}[!t]
\begin{center}
\includegraphics[width=.47\textwidth]{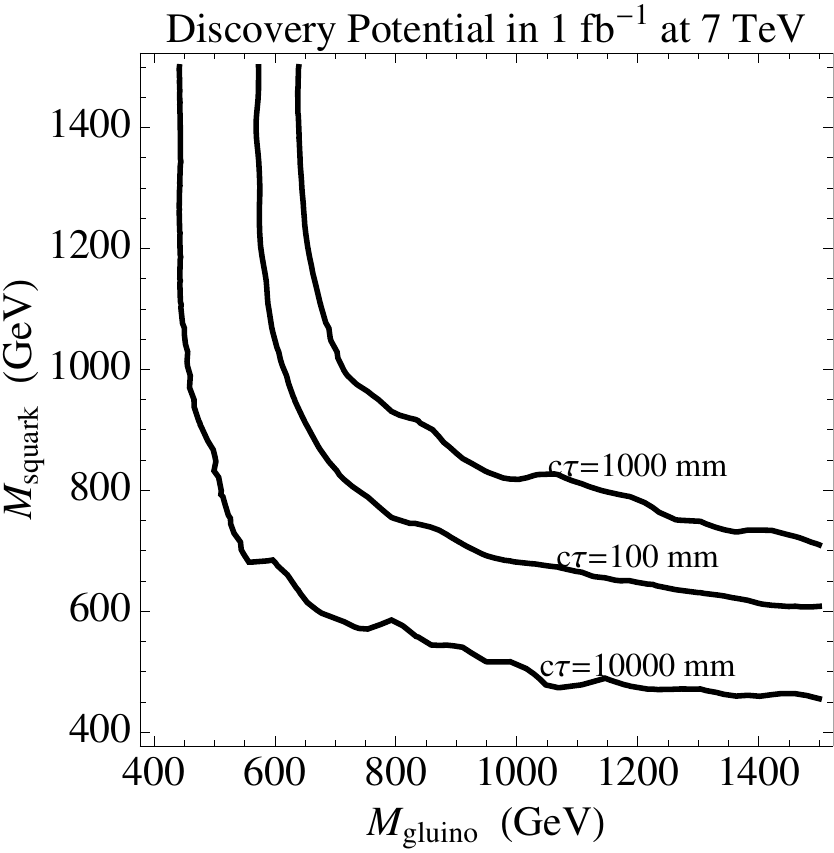}\qquad \includegraphics[width=.47\textwidth]{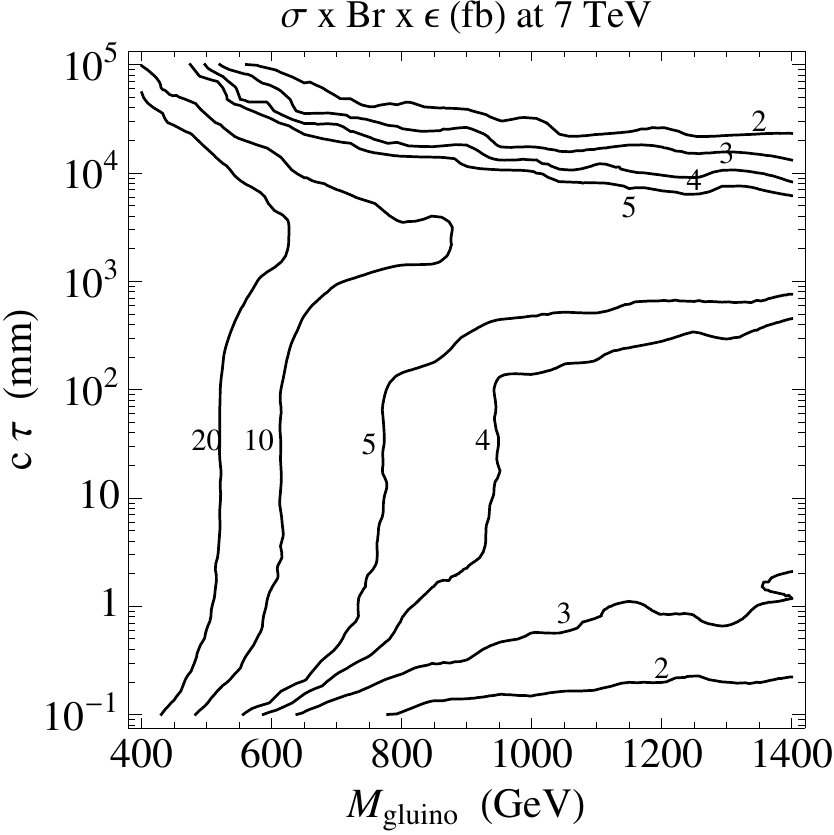}
\end{center}
\caption{Discovery potential for Higgsino NLSPs decaying to displaced $Z$'s  (in the final states discussed in this section) at ATLAS in 1 fb$^{-1}$ at 7 TeV.  Left plot: lines of constant $\sigma\times{\rm Br}\times\varepsilon=5$, for different values of $c\tau$, in the $M_{gluino}$-$M_{squark}$ plane. Right plot:  Contours of $\sigma\times{\rm Br}\times\varepsilon$ in the $M_{gluino}$-$c\tau$ plane, for fixed $M_{squark}=1$ TeV. In both plots we have taken $m_{NLSP}=250$ GeV.
}
\label{fig:xsecdisc}
\end{figure}

As mentioned at the beginning of this section, at the LHC we imagine that production of colored sparticles (gluinos, squarks) is the primary source of neutralino NLSPs. To estimate the discovery reach in the lifetime and colored cross section plane, we take the following benchmark scenario: gluinos and squarks decaying directly down to Higgsino NLSPs.  We will fix the NLSP mass at 250 GeV, since the discovery potential does not depend strongly on it in most of the parameter space. We also assume for simplicity that $Br(\tilde{\chi}_1^0\to Z+\tilde G)=1$.    We expect that the discovery potential is insensitive to the details of the spectrum between the colored sparticles and the NLSP, because the analyses described above are fully inclusive.

Varying the gluino, common squark mass, and lifetime, we have calculated $N_{pass}$, the number of events in 1 fb$^{-1}$ at 7 TeV passing an OR of all the analyses described above. Shown on the left in fig.\ \ref{fig:xsecdisc} are contours of constant $N_{pass}=5$ in the $M_{squark}$, $M_{gluino}$ plane, for different values of the lifetime. On the right in fig.\ \ref{fig:xsecdisc} are contours of constant $N_{pass}$ in the $M_{gluino}$, $c\tau$ plane for $M_{squark}=1$ TeV.   Note that in the right panel of Figure~\ref{fig:xsecdisc} there are bands of constant $N_{pass}$ as a function of the gluino mass, centered around $c\tau\sim 1$ m.  These bands exist because there is still EW production of charginos and neutralinos at the LHC.  Even for a modestly heavy Higgsino NLSP, without having colored particles within reach, the LHC at 7 TeV {\em still} has discovery potential during the first inverse fb.
With the inclusion of colored sparticles, the search for displaced $Z$ decays  becomes a golden channel for the discovery of new physics in the first year of the LHC.

%%%%%%%%%%%%%%%%%%%%%%%%%%%%%%%%%%%%%%%%%%%%%%%%%%%
\section{Detailed Event Study at 14 TeV}
\label{sec:atlasrecon}
\setcounter{equation}{0} \setcounter{footnote}{0}
%%%%%%%%%%%%%%%%%%%%%%%%%%%%%%%%%%%%%%%%%%%%%%%%%%%

Having explored the discovery potential in early LHC data, we will now turn to detailed, post-discovery analysis of displaced $Z$ decays at the LHC design energy (14 TeV). We will show that by combining the capabilities of different parts of the ATLAS detector, full kinematic reconstruction of the event is possible. That in turn will enable detailed properties of the NLSP to be measured precisely, including its mass, lifetime and wino/Higgsino  content.

\subsection{Sample Points}

Using Herwig 6.5 with full spin correlations \cite{Herwig}, we have simulated some sample points to test the ability to fully reconstruct events at 14 TeV. The parameter choices, corresponding total SUSY production cross sections, and cross sections for events containing the $\tilde\chi^0_1 \to \tilde G~Z(\to e^+ e^-)$ are listed in Table \ref{tab:points}. We have chosen points with two NLSP masses, $\approx 250$ GeV and $\approx 450$ GeV (denoted ``L" and ``H"); two gauge eigenstate contents, Higgsino and wino (denoted ``H" and ``W"); and two gluino masses, 1000 GeV and 600 GeV (the latter denoted with an ``S"). For the most part we will show plots for the optimistic point LHS with a 250 GeV Higgsino and a 600 GeV gluino, but some more detailed comparisons of the points appear in Table \ref{tab:pointsresults} and Section \ref{sec:angulardists}. For the Higgsino points, roughly half of the $\tilde{\chi}^0_1$ decays are to $Z$; this could be increased to nearly 100\% by taking $\tan \beta \approx 1$.

\begin{table}[t!]
\centering
\begin{tabular}{|c|c|c|c|c|c|c|}
\hline
Point & $\mu$ & $M_2$ & $M_{\tilde g}$ & $\sigma_{\rm tot}$ & $\sigma_{Z\to e^+ e^-}$ \\
 \hline
LH & 250 & 800 & 1000 & 2.0 pb & 79 fb \\ % 10293 gen events = 130 fb^-1
LHS & 250 & 800 & 600 & 12.9 pb & 506 fb \\ % 295138 gen events = 583 fb^-1
LW & 800 & 245 & 1000 & 2.0 pb & 30 fb \\ % 8850 gen events = 295 fb^-1
LWS & 800 & 245 & 600 & 12.4 pb & 200 fb\\ % 116839 gen events = 584 fb^-1
\hline
HH & 450 & 800 & 1000 & 1.8 pb & 47 fb \\ % 4958 gen events = 105 fb^-1
HHS & 450 & 800 & 600 & 12.2 pb & 388 fb \\ % 238071 gen events = 614 fb^-1 
HW & 800 & 445 & 1000 & 1.9 pb & 34 fb \\ % 12044 gen events = 354 fb^-1
HWS & 800 & 445 & 600 & 12.2 pb & 233 fb \\ % 143499 gen events = 616 fb^-1
\hline
\end{tabular}
\caption{Simulated points. Designations: first letter is ``L" or ``H" for ``light" or ``heavy" neutralino NLSP (242 GeV and 438 GeV, respectively); second letter is ``H" or ``W" for ``Higgsino-like" or ``wino-like" NLSP; ``S" as the third letter means ``enhanced strong production", i.e. a relatively light gluino. Parameters not listed in the table are identical for every point: $M_1 = 800$ GeV all squarks and sleptons are at 1 TeV, and $\tan \beta = 20$. Cross sections are as reported by Herwig 6.5 \cite{Herwig}.}
\label{tab:points}
\end{table}

\subsection{Cuts, Acceptance, and Efficiency}
\label{sec:acceffdef}

\subsubsection{ECAL and TRT}
\label{sec:accdefecal}

We impose a set of cuts that is intended to give a rough characterization of which events are potentially reconstructible and low-background. These cuts are listed in Table \ref{table:accee}. If only the ECAL is used for reconstruction, the cut on detector eta is weakened to $|\eta_{det}| < 1.4$, as the electrons are no longer required to pass through the TRT. 

We have included here two cuts which were not used in the discovery analysis of section \ref{sec:discovery}. They arise because pointing and timing information are necessary in order to fully reconstruct the events. First, we require that the effective $z$-vertex satisfies $|z_{e.v.}| < 120~{\rm cm}$, so that the pointing angle resolution $\sigma_\theta$ remains reasonably small \cite{Prieur05}. Second, to guarantee that we look at events that are noticeably delayed in time, we require a time delay (relative to an electron from the primary vertex) of 0.3 ns for both electrons.  In simulating the time delay we assume that the beam spot is essentially Gaussian with $\sigma_z = 5.6$ cm ($\sigma_{x,y} = 15 \mu{\rm m}$). We assume that the primary vertex $z$ position will be measured well enough that smearing by a Gaussian of width 100 $\mu$m gives a reasonable description of its uncertainty \cite{ATLASTDRv1}. We will refer to the fraction of events with a $Z \to e^+ e^-$ decay that pass these cuts as the ``acceptance."

\subsubsection{Muons}
\label{sec:accdefmu}

Because the muon system does not have separate tracking and pointing, but simply measures 3d tracks and their arrival time, and because the acceptance cuts in section \ref{sec:discovery} already included a requirement of large decay radius $r_d > 500$ mm, we use the same acceptance cuts as in the discovery analysis. They are listed in Table \ref{table:accmumu}.

\subsubsection{Reconstruction}
\label{sec:accdefrecon}

With either the ECAL+TRT combination or the muon spectrometer alone, the kinematics of the decay are {\em overconstrained} (12 measurements for 10 unknowns). To reconstruct the events, we begin with a ``seed choice." The goal is to take the measured parameters for each electron, $(E, \eta_{det}, \varphi_{det}, \theta_{dir}, \varphi_{dir}, t_{det})$ and produce a preliminary description of the event kinematics, as specified by $(x_d, y_d, z_d, t_d)$ and ${\bf p}(e^{\pm})$. We determine $(x_d, y_d)$ by finding the intersection of the lines in the $r-\varphi$ plane determined by $(\varphi_{det}, \varphi_{dir})$ for the two electrons. Given $(x_d, y_d)$, we then estimate $(z_d, t_d)$ by choosing one of the electrons (arbitrarily) and extrapolating backward (using its arrival time $t_d$ and angle $\theta_{dir}$). This gives an initial seed, from which we can then search for a local minimum of the $\chi^2$ for the measured quantities (i.e. $\sum_i (O_i - E_i)^2/\sigma_i^2$). The resolutions $\sigma_i$ depend on measured quantities like $E$ and $z_{e.v.}$; we compute these from the initial seed point, rather than allowing $\sigma$ to vary across the set of values we minimize over.

We consider the event ``well-reconstructed" if it additionally satisfies a tight $Z$ mass window cut 81 $< m_Z < 101$ GeV and a loose NLSP mass window cut 90 $< m_{\tilde{\chi}^0_1} < 1000$ GeV. The fraction of events that are well-reconstructed by this criterion will be referred to as the ``efficiency." In practice, the acceptance and efficiency are similar, indicating that our acceptance cuts select reconstructible events almost always.

\begin{figure}[!t]
\begin{center}
\includegraphics[width=0.5\textwidth]{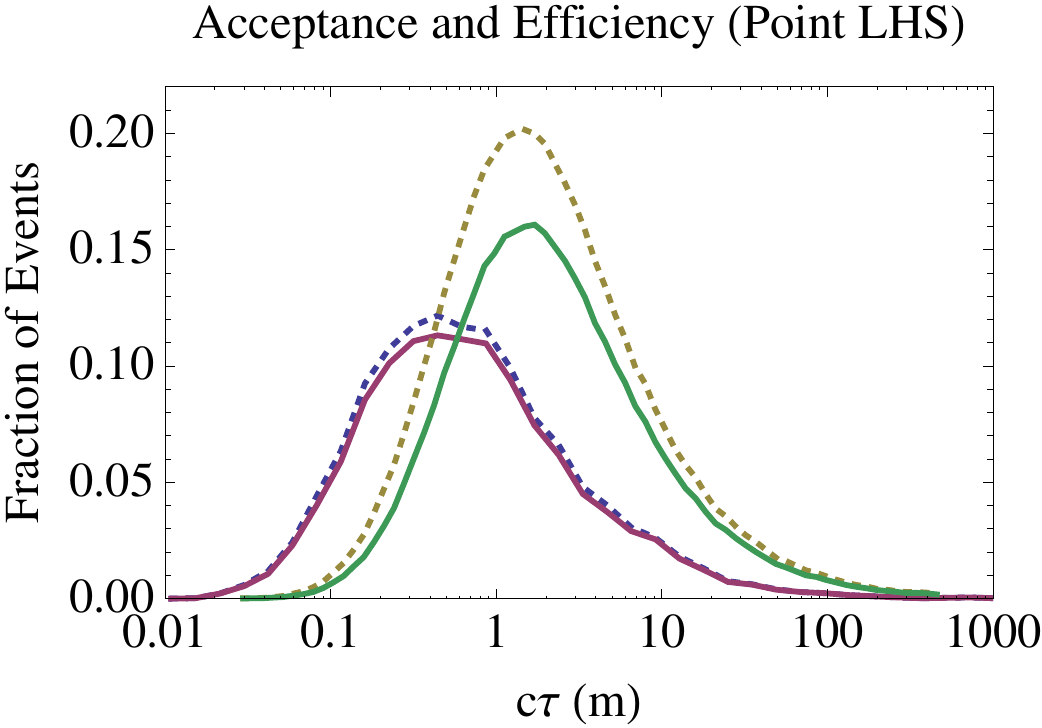}
\end{center}
\caption{Geometric and kinematic acceptance (dashed blue line for ECAL+TRT, dashed yellow line for muons) and reconstruction efficiency (solid purple line for ECAL+TRT, solid green line for muons) as a function of lifetime for point LHS. The reconstruction efficiency is defined as the fraction of {\em all} $Z \to \ell^+ \ell^-$ events that are well-reconstructed.}
\label{fig:acceff}
\end{figure}

In Figure \ref{fig:acceff}, we show how the acceptance and efficiency change as a function of lifetime.  At $c\tau$ of about 1 meter, the muon spectrometer takes over from the ECAL as the preferred way to measure long-lived decays. At very long lifetimes, the trigger requirement of a time delay below 6 ns is somewhat limiting, but the muon spectrometer still has sensitivity to lifetimes beyond 100 meters provided cross sections are not too small. To give a sense of how efficiencies and mass reconstruction vary as a function of the spectrum, we fix a lifetime of 1 ns (so that the ECAL is the preferred detector component) and tabulate efficiencies and reconstructed masses for all of the samples in Table \ref{tab:pointsresults}. We also include the results of the ECAL-only reconstruction for comparison. It gives comparable, or sometimes higher, efficiencies, but broader and less accurate distributions.

\begin{table}[t!]
\centering
\begin{tabular}{|c||c|c|c|c||c|c|c|c|}
\hline
Point & $m_{rec}$ & $\sigma_{m_{rec}}$ & $N_{ev}$ & $\epsilon$ & $m^{(ECAL)}_{rec}$ & $\sigma^{(ECAL)}_{m_{rec}}$ & $N^{(ECAL)}_{ev}$ & $\epsilon^{(ECAL)}$ \\
 \hline
LH & 242.9 & 1.8 & 61.9 & 0.078 & 239.5 & 4.5 & 75.4 & 0.095 \\
LHS & 242.6 & 0.74 & 571 & 0.113 & 240.2 & 0.92 & 791 & 0.156  \\
LW & 242.3 & 4.0 & 14.5 & 0.048 & 238.2 & 6.5 & 22 & 0.073  \\
LWS & 242.1 & 1.3 & 193 & 0.096 & 241.7 & 1.5 & 268 & 0.134 \\
\hline
HH & 436.4 & 2.4 & 58.3 & 0.124 & 428.2 & 12.6 & 76.7 & 0.163\\
HHS & 436.2 & 0.92 & 546 & 0.141 & 424.9 & 3.8 & 678 & 0.175 \\
HW & 435.5 & 4.8 & 37.9 & 0.111 & 433.4 & 18.4 & 45.8 & 0.135 \\
HWS & 436.8 & 1.1 & 293 & 0.126 & 428.9 & 4.6 & 359 & 0.154 \\
\hline
\end{tabular}
\caption{Ability to reconstruct the various simulation points; for each, ten independent 10 fb$^{-1}$ samples were run with the lifetime fixed at 1 ns. We report the average and standard deviation of the reconstructed mass $m_{rec}$ (found by Gaussian fit seeded with the median of distribution) over the ten samples, and the average number of reconstructed events $N_{ev}$ and corresponding efficiency $\epsilon$. The final four columns report results relying only on the ECAL.}
\label{tab:pointsresults}
\end{table}

\subsection{Mass reconstruction}

\begin{figure}[!t]
\begin{center}
\includegraphics[width=0.45\textwidth]{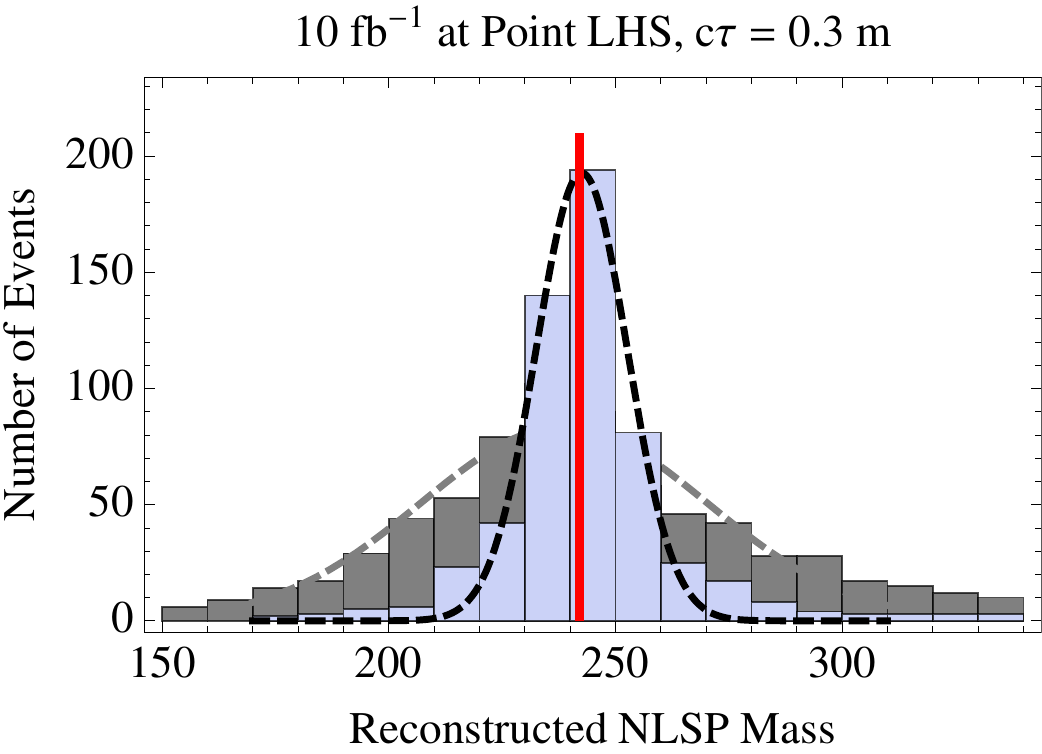}~\includegraphics[width=0.45\textwidth]{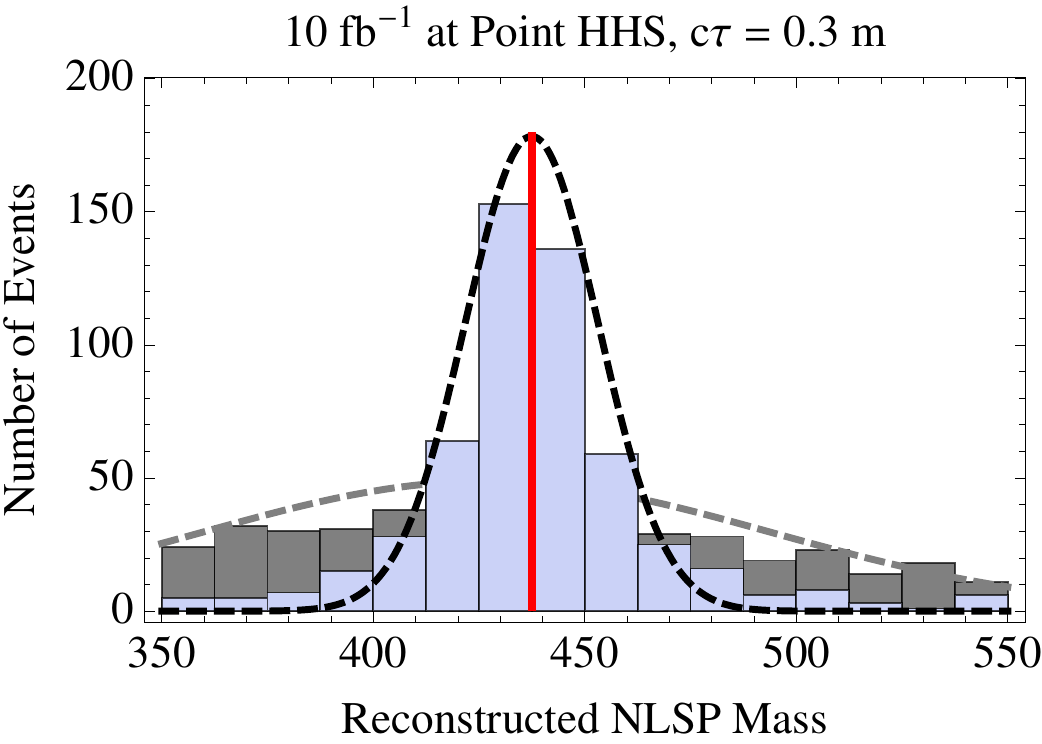}
\end{center}
\caption{Reconstructed $m_{\tilde{\chi}^0_1}$ for a Higgsino-like NLSP, using the ECAL and the TRT. Here we have restricted to events with $m_Z$ reconstructed within 10 GeV of the true value, and performed a nonlinear least-squares fit to a Gaussian. The points are LHS and HHS, and the Gaussian fit is accurate within 1 GeV for both. The gray histograms and dotted lines in the background show the ECAL-only reconstruction and corresponding Gaussian fit.}
\label{fig:ecaltrtfit}
\end{figure}

In Figure \ref{fig:ecaltrtfit}, we show histograms of reconstructed neutralino masses in a 10 fb$^{-1}$ sample of events passing the cuts of section \ref{sec:acceffdef}, together with a nonlinear least-squares fit to a Gaussian to find the mass. For the point LHS, there are 554 well-reconstructed events; for the point HHS, there are 561. In both cases, we also show, in the background, the results of a fit using only information from the ECAL. Notice that the ECAL-only efficiency is comparable, as a similar number of events are reconstructed, but the resolution is noticeably worse, and a typical event will be reconstructed far more accurately if the TRT is used. This highlights the importance of understanding how to construct standalone TRT tracks originating at a given calorimeter cell that do not necessarily point near the primary vertex.

\begin{figure}[!t]
\begin{center}
\includegraphics[width=0.45\textwidth]{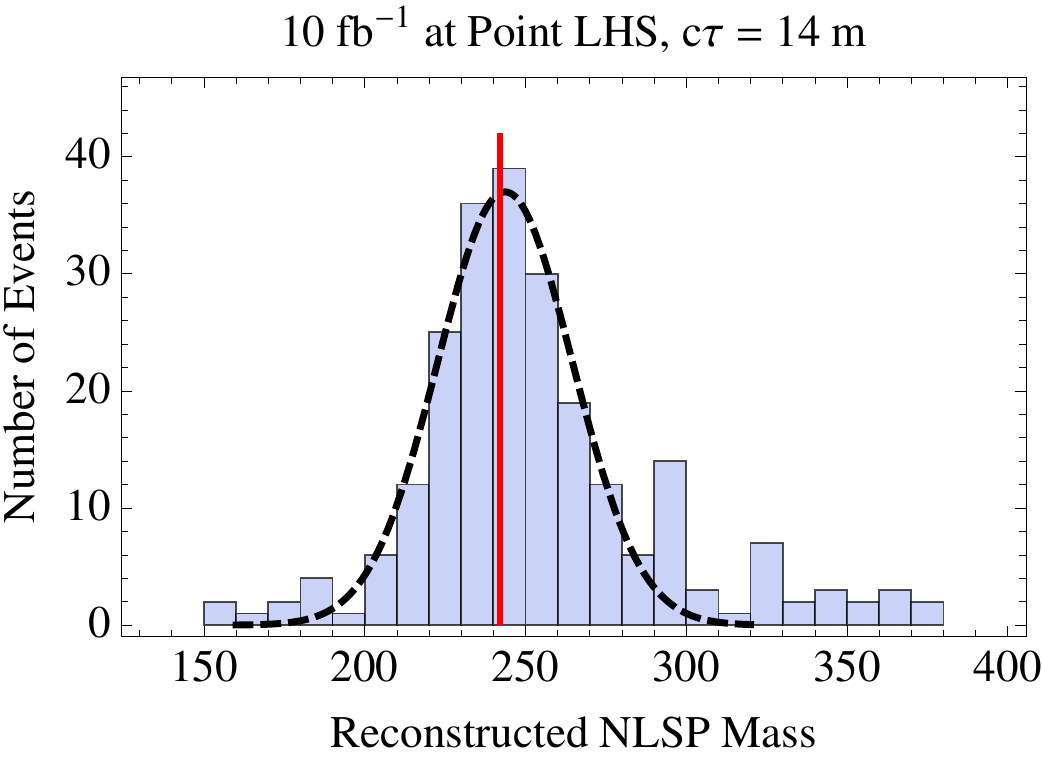}~\includegraphics[width=0.45\textwidth]{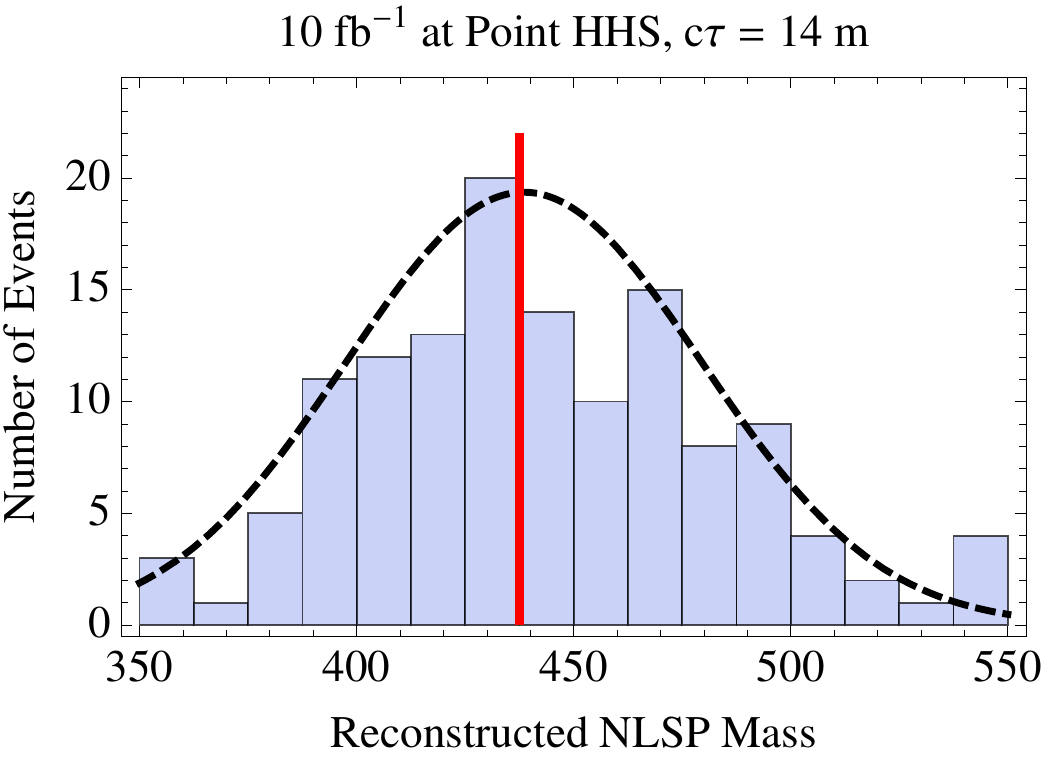}
\end{center}
\caption{Reconstructed $m_{\tilde{\chi}^0_1}$ for a Higgsino-like NLSP, in points LHS and HHS, using the muon spectrometer. This demonstrates the capacity to reach significantly longer lifetimes.}
\label{fig:mufit}
\end{figure}

Figure \ref{fig:mufit} shows a similar result for reconstruction using the muon spectrometer. Here the lifetime is 14 m, a factor of 40 larger than that shown in Figure \ref{fig:ecaltrtfit}. For this lifetime there are 286 well-reconstructed events at point LHS and 150 at point HHS.

\subsection{Lifetime Measurement}
\label{sec:lifetime}

\begin{figure}[!t]
\begin{center}
\includegraphics[width=0.45\textwidth]{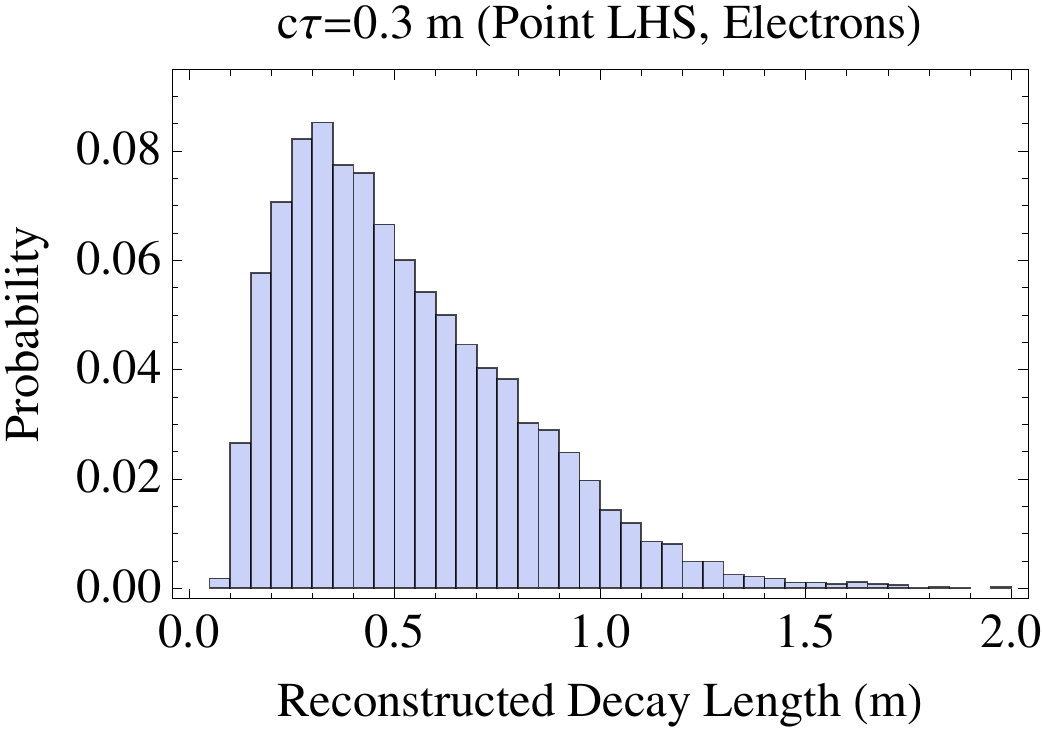}~~\includegraphics[width=0.45\textwidth]{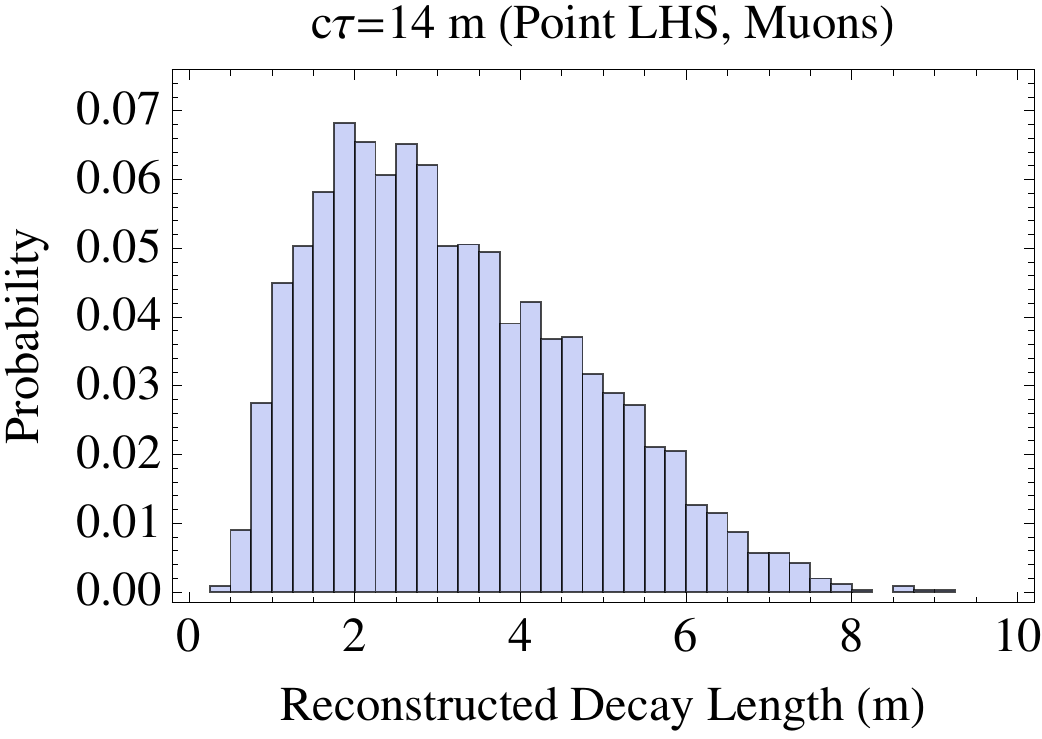}
\end{center}
\caption{Distribution of reconstructed decay lengths, for a case using the ECAL and TRT (at left, $c\tau = 0.3$ m) and using the muon spectrometer (at right, $c\tau = 14$ m).}
\label{fig:reconlengths}
\end{figure}

In GMSB, one of the most interesting quantities to measure is the lifetime of the NLSP, from which the fundamental scale of SUSY breaking $\sqrt{F}$ can be derived. Our cuts require delays of 0.3 ns for reconstruction with the ECAL, and decay radii of 500 mm for use of the muon spectrometer, so finding a sample of well-reconstructed events with a clean mass peak automatically implies that one is viewing events with a long lifetime. However, the distribution of reconstructed lifetimes is shaped by multiple effects. First, the distribution of proper lifetimes $e^{-t/\tau}$ is convolved with the effects of boost factors which depend on the mass of the {\em produced} particles (like gluinos or squarks) as well as the NLSP mass. On an event-by-event basis, the boost can be measured and corrected for, but with some error. Second, the geometry of the detector imposes limitations; for instance, for events reconstructed with the ECAL, decays at $t \gappeq 3$ ns will be too long-lived to be detected, so the tail of the distribution is chopped off. Events at lifetimes that are too {\em short} are not reconstructible, so the beginning of the distribution is also shaped by cuts. As a result, very little remnant of an exponential distribution of proper times is left (see Figure \ref{fig:reconlengths}).

The unknown spectrum presents a challenge, but given the good kinematic reconstruction of decays that we have shown is achievable, it seems likely that masses of squarks and gluinos can be measured well. The cross section also contains information about these masses. Such mass determination problems are orthogonal to the subject of this paper, so we will not consider them in detail here. For the time being, we will simply assume that the unknown spectrum has been independently measured, so that the distribution of NLSP lifetimes can be simulated accurately.

\begin{figure}[!h]
\begin{center}
\includegraphics[width=0.5\textwidth]{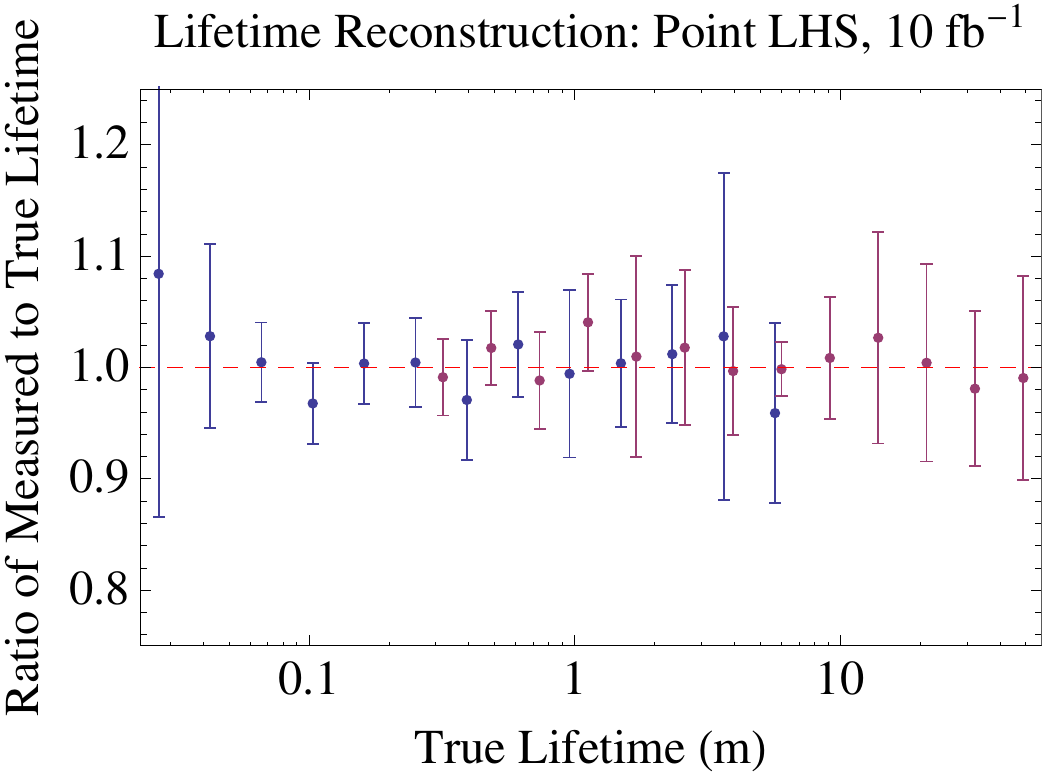}
\end{center}
\caption{Ability to reconstruct lifetimes: mean and standard deviation of the ratio of measured to true lifetime inferred from 10 independent pseudoexperiments of 10 fb$^{-1}$ each, by $\chi^2$ fitting of the distribution of reconstructed times for each event. We show the point LHS, which has a large cross section, and for which the lifetime can be measured accurately. Blue points are measured with electrons in the ECAL+TRT, whereas purple points are measured with the muon spectrometer.}
\label{fig:timerecon}
\end{figure}

Assuming that the spectrum has been measured, one can simulate a variety of different lifetimes and perform a $\chi^2$ fit to find the true lifetime. We show the results of such a fit in Figure \ref{fig:timerecon}. A wide range of lifetimes can be measured with $\sim 10\%$ accuracy. It is likely that one of the best tools for understanding lifetime is the ratio of events measured in different parts of the detector, as we discussed in section \ref{sec:geom}, provided that the efficiencies and kinematics are understood. For relatively short lifetimes, it would interesting to consider how well the silicon can be used to measure $c\tau$, although such events would not be fully reconstructible as they would occur at times to short to use the ECAL timing system.

\subsection{NLSP Angular Distributions}
\label{sec:angulardists}

So far we have focused on reconstructing basic kinematic information like masses and decay widths. More detailed information can be obtained by studying angular distributions. For example, a neutral wino decays to transversely polarized $Z$ bosons, whereas a neutral Higgsino decays to gravitino plus longitudinal $Z$. In the decay chain $\tilde\chi^0_1 \to \tilde G Z(\to e^+ e^-)$, the invariant mass 
\begin{equation}
M^2(e^-,\tilde G) = \frac{1}{2} \left(m_{\tilde \chi^0_1}^2 - m_Z^2\right) (1 + \cos \theta_*)
\end{equation}
where $\theta_*$ is the angle in the $Z$ rest frame between the $e^-$ direction and the boost direction from the $\tilde\chi^0_1$ rest frame to the $Z$ rest frame. In this variable, the expected angular distribution is $1 + \cos^2 \theta_*$ for transversely polarized Z bosons and $1 - \cos^2 \theta_*$ for longitudinally polarized Z bosons. Similar techniques were discussed in the context of the hypothetical ILC in \cite{Ambrosanio:1999iu}. Note that this is an idealized theoretical prediction that will be shaped by cuts. (In particular, $\cos \theta_* \to 1$ is a limit in which the $e^+$ is soft and the event will fail our cuts.)

\begin{figure}[!h]
\begin{center}
\includegraphics[width=0.46\textwidth]{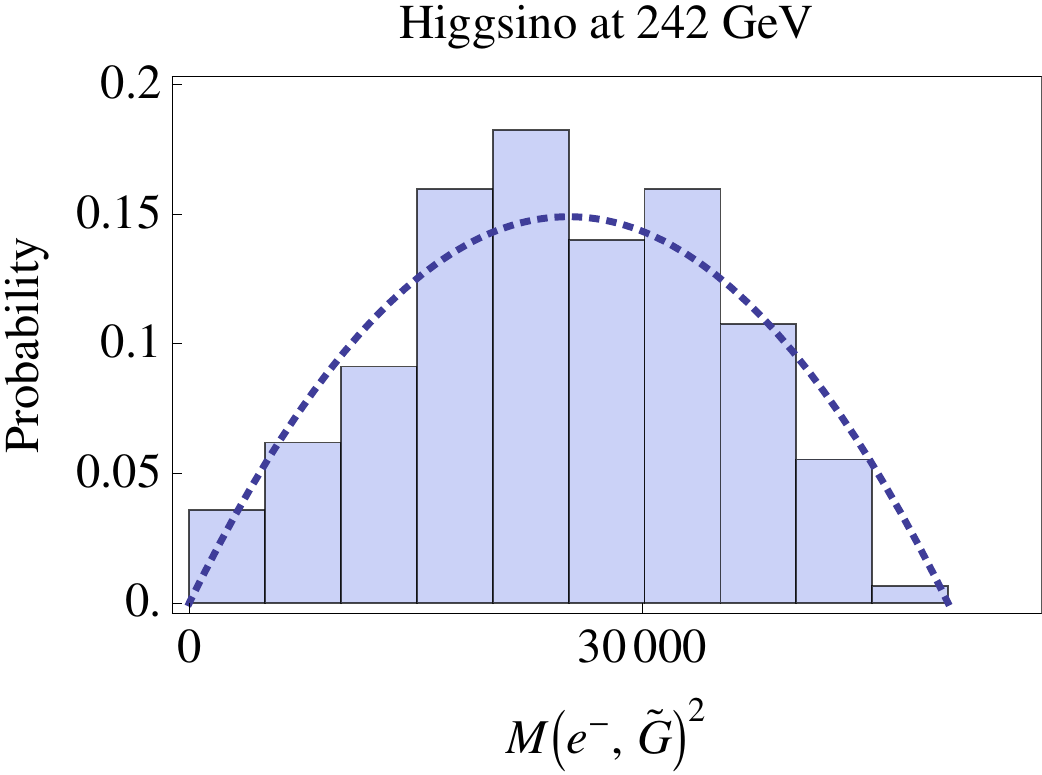}~\includegraphics[width=0.46\textwidth]{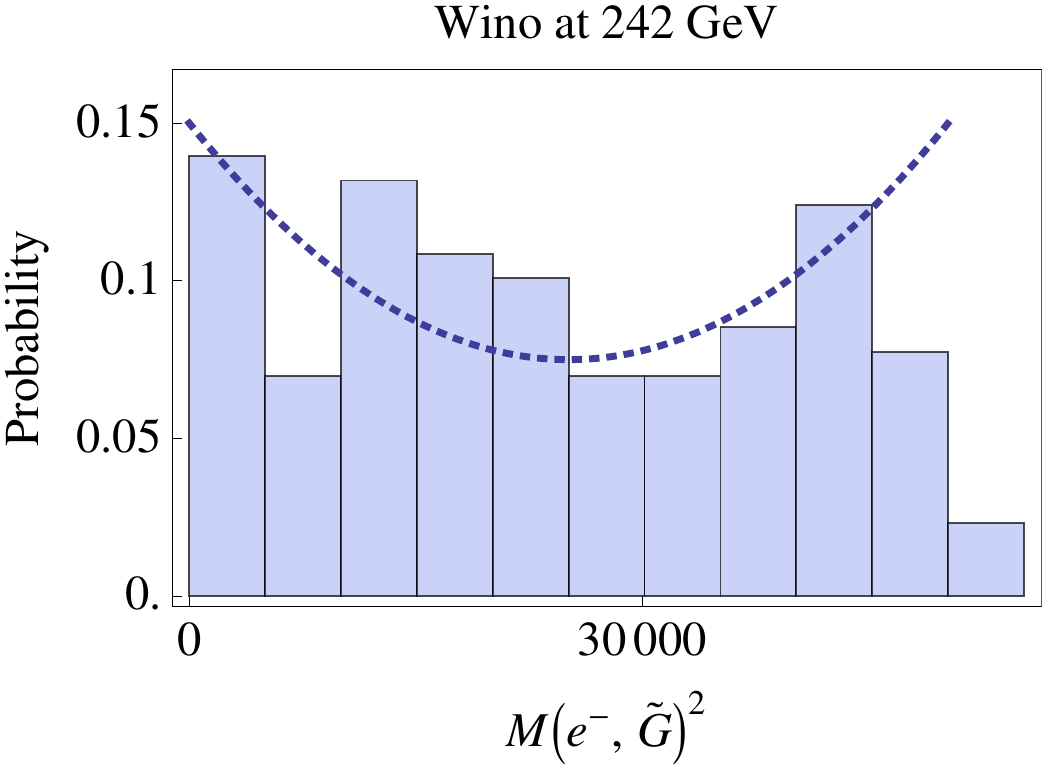}
\end{center}
\caption{Invariant mass squared $M^2(e^-,\tilde G)$ of the electron and gravitino in a neutralino decay chain, after reconstruction of the event, with 10 fb$^{-1}$ of data and $c\tau = 0.3$ m. At left: a mostly-Higgsino NLSP (model point LHS), with dotted line showing the idealized longitudinal distribution $\propto 1 - \cos^2\theta_*$. At right: a mostly-wino NLSP (model point LWS), with dotted line showing the idealized transverse distribution $\propto 1 + \cos^2\theta_*$.}
\label{fig:HWcompare}
\end{figure}

In Figure \ref{fig:HWcompare} we illustrate the ability of ATLAS to discriminate between transversely and longitudinally polarized Z bosons (and, hence, neutral wino vs Higgsino NLSPs). A Kolmogorov-Smirnov test shows that these distributions differ at the 99\% confidence level. This is based on points LHS and LWS, i.e. two points with a 600 GeV gluino to enhance the production cross section. The lifetime is 1 ns and we have used ECAL and TRT information in the reconstruction. In addition to the cuts we took to define a ``well-reconstructed event" in section \ref{sec:accdefrecon}, we further keep only events for which the reconstructed value of $m_{\tilde{\chi}^0_1}$ is within 10 GeV of the best-fit value, which gives a cleaner sample. This simple example illustrates that we can move beyond studying just masses and branching ratios to actually probing the structure of the Lagrangian.

%%%%%%%%%%%%%%%%%%%%%%%%%%%%%%%%%%%%%%%%%%%%%%%%%%%
\section{Discussion}
\label{sec:discussion}
\setcounter{equation}{0} \setcounter{footnote}{0}
%%%%%%%%%%%%%%%%%%%%%%%%%%%%%%%%%%%%%%%%%%%%%%%%%%%

In this paper we have established that the LHC can discover long-lived neutral particles decaying to $Z$ bosons, with lifetimes ranging from about 0.1 mm to 100 meters, using just 1 fb$^{-1}$ of data at 7 TeV. We have seen that this is a clean, well-motivated, and powerful discovery channel for new physics, and it is one that has up till now been neglected at the LHC. Thus it presents an exciting prospect that could be put to the test within the next two years. 

Because our long-lived particles decay to multiple charged particles, and these particles arrive with a time delay in a direction that does not point back to the beamline, the problem of reconstructing the event kinematics becomes highly overconstrained. 
For lifetimes from a few centimeters to tens of meters, we have shown that the ECAL and TRT or the muon spectrometer provide enough precision information to allow masses, lifetimes, and even angular distributions to be accurately measured with 10 fb$^{-1}$ of data at 14 TeV. This should be contrasted with prompt decays or even displaced decays to photons, where much less information is available, and one must work much harder to reconstruct these quantities.

We have not necessarily exhausted the full set of options for using the ATLAS detector to study long lifetimes; for instance, the Tile (hadronic) calorimeter can measure arrival time with resolution ranging from hundreds of picoseconds to 1.5 ns depending on energy \cite{TilecalTiming}. In general, we expect that decays inside a calorimeter will offer less precise directional information than those in a tracking system, but they can still be interesting both for discovery and to get a better understanding of lifetime by counting decays in different detector components.

Our results have been specialized to ATLAS, but most of them carry over to CMS with minimal changes.  Perhaps the most important difference is simply the volume of the two detectors: CMS has a radius of 7.3 meters \cite{CMSTDR}, whereas the muon spectrometer at ATLAS extends out to 10 meters. Thus CMS can be expected to have a somewhat more limited ability to probe very long lifetimes. Aside from the geometry, however, CMS should perform well in measuring standalone muons from long-lifetime decays, with full 3d directional information, and a timing resolution of $\sim 1$ ns from RPCs \cite{CMSRPC}. The CMS detector also has an EM timing system, which like that of ATLAS has a resolution of 100 ps for EM showers (and has achieved synchronization across crystals of 500 ps with early events) \cite{CMSEMTiming}. The ability of the CMS ECAL to measure non-pointing angles is more limited, but at a crude level it is possible \cite{CMSnonpointing,CMSGMSB}. On the other hand, whereas the ATLAS TRT provides information only in the $r-\varphi$ plane, at CMS the first two layers of both the TIB (``Tracker Inner Barrel") and TOB (``Tracker Outer Barrel") are stereo \cite{CMSTDR}, so provided a decay is prompt enough, the tracker can be used to provide 3d directional information in place of pointing information from the calorimeter. The stochastic term in the CMS ECAL is quite good, $\sigma_E \approx 0.036 \sqrt{E~{\rm GeV}}$, although as with standard photon or electron measurements at CMS, the large amount of material in the tracker could complicate precise measurement.

It is interesting to consider to what extent the reconstruction techniques we discuss can be pushed to shorter lifetimes. It is conceivable that future detector upgrades could have picosecond-resolution timing \cite{Genat:2008vc}, giving two orders of magnitude improvement and allowing silicon to be used in place of pointing for 3d directional information. The use of such high-precision timing would essentially allow the entire discovery reach discussed in section \ref{sec:discovery} to be accurately reconstructed along the lines of section \ref{sec:atlasrecon}.

%%%%%%%%%%%%%%%%%%%%%%%%%%
\section*{Acknowledgments}
We have benefited from conversations with many experimentalists: we thank Andrew Askew, Adam Aurisano, Yuri Gershtein, Max Goncharov, Beate Heinemann, John Hobbs, Tom LeCompte, Jason Nielsen, Damien Prieur, Evelyn Thomson, Ian Tomalin, Dan Ventura, and Dirk Zerwas for helpful comments and informative responses. We also thank Matt Strassler, Scott Thomas, and Lian-Tao Wang for useful discussions. We thank the Aspen Center for Physics for hospitality while the work was in progress. P.M. and M.R. thank the Kavli Institute for Theoretical Physics in Santa Barbara, CA for hospitality while this work was initiated. M.R. would like to thank the participants and organizers of the May 2009 workshop on long-lived particles at UW Seattle (supported by the DOE under Task TeV of contract DE-FGO2-96-ER40956) for informative and provocative discussions, and the PCTS for its support. The work of PM was supported in part by NSF grant PHY-0653354.  The work of DS was supported in part by DOE
grant DE-FG02-90ER40542 and the William D. Loughlin membership at
the Institute for Advanced Study.  

%%%%%%%%%%%%%%%%%%%%%%%%%%%%%%%%%%%%%%%%%%%%%%%%%%%
\appendix

%%%%%%%%%%%%%%%%%%%%%%%%%%%%%%%%%%%%%%%%%%%%%%%%%%%
\section{Tevatron Searches for Long-Lived Neutral Particles}
\label{sec:tevatron}
\setcounter{equation}{0} \setcounter{footnote}{0}
%%%%%%%%%%%%%%%%%%%%%%%%%%%%%%%%%%%%%%%%%%%%%%%%%%%

In this appendix, we will discuss Tevatron searches for delayed decays of long-lived, heavy, neutral particles~\cite{cdfgammajetmettiming, longzeecdfrun1, longlivezmumucdf, longmumud0, longzeeggd0, longlivebbd0}. We will focus on their discovery potential for general neutralino NLSPs decaying to $Z(\ell^+\ell^-)+\met$. Currently, there are no Tevatron analyses which directly search in this final state. However, several searches motivated by other theoretical considerations have the potential to indirectly constrain this scenario.

CDF has performed two searches for a long-lived neutral particle decaying to a $Z$ boson and possibly other particles, visible or invisible. These studies use well-reconstructed tracks found by the default algorithms, so in particular, any particle that decays at a radius larger than $\sim 1$ cm would be missed.  Both were motivated by a possible fourth generation quark $b' \to bZ$ decaying through a loop. The first searched for $Z \to e^+ e^-$ in a 90 pb$^{-1}$ sample \cite{longzeecdfrun1}, whereas the second searched for $Z \to \mu^+ \mu^-$ in a 163 pb$^{-1}$ sample \cite{longlivezmumucdf}. Both studies imposed a $Z$ mass window cut and required a minimum $L_{xy}$, the distance in the transverse plane between the interaction point and the vertex where the lepton pair originates, of order 0.1 cm. Both searches set limits of order 1 pb for lifetimes near 0.1 to 1 cm. 

\dzero has carried out studies that offer more flexibility in the range of lifetimes probed, by not requiring standard track quality cuts. One study used 380 pb$^{-1}$ of data to search for a displaced $\mu^+\mu^-$ pair \cite{longmumud0}, motivated by the NuTeV anomaly and RPV decays of a neutralino. The muons were required to have a minimum distance of closest approach of the tracks to any vertex in both the transverse plane (0.01 cm) and the $z$ direction (0.1 cm), and were {\em not} required to leave silicon hits. This search required $\Delta R(\mu^+ \mu^-) < 0.5$, which leads to its optimal reach being at lower masses than the $Z$, so it is not optimal for gauge mediation searches. The limit was about 0.1 pb for a lifetime of about 1 cm.

More interestingly for us, \dzero has undertaken a study with 1.1 fb$^{-1}$ of data that searches for displaced electrons or photons using pointing~\cite{longzeeggd0} and not using tracking. The pointing measurement gives full 3d information on the direction of an object interacting with the EM calorimeter, by fitting a line through five shower positions. This allows a great deal more flexibility in the range of lifetimes probed than studies that rely on tracking information. This is the only Tevatron search that we are aware of which constrains this scenario using timing or pointing. This search is sensitive to longer displacements ($\sim\mathcal{O}(10-100)$ cm) than the tracking-based searches, and so it offers a complementary approach to the searches described above. 

Since this search was again motivated at least partly by the $b'\to b Z$ scenario, the cuts were not optimized for gauge mediation. In particular, there was no requirement on missing energy, and consequently there was a nonzero expected background. We expect that with a hard MET cut, the background  could have been completely eliminated, so that with more data, this channel could become a clear discovery mode for GMSB.

\begin{figure}[!t]
\begin{center}
\includegraphics[width=0.6\textwidth]{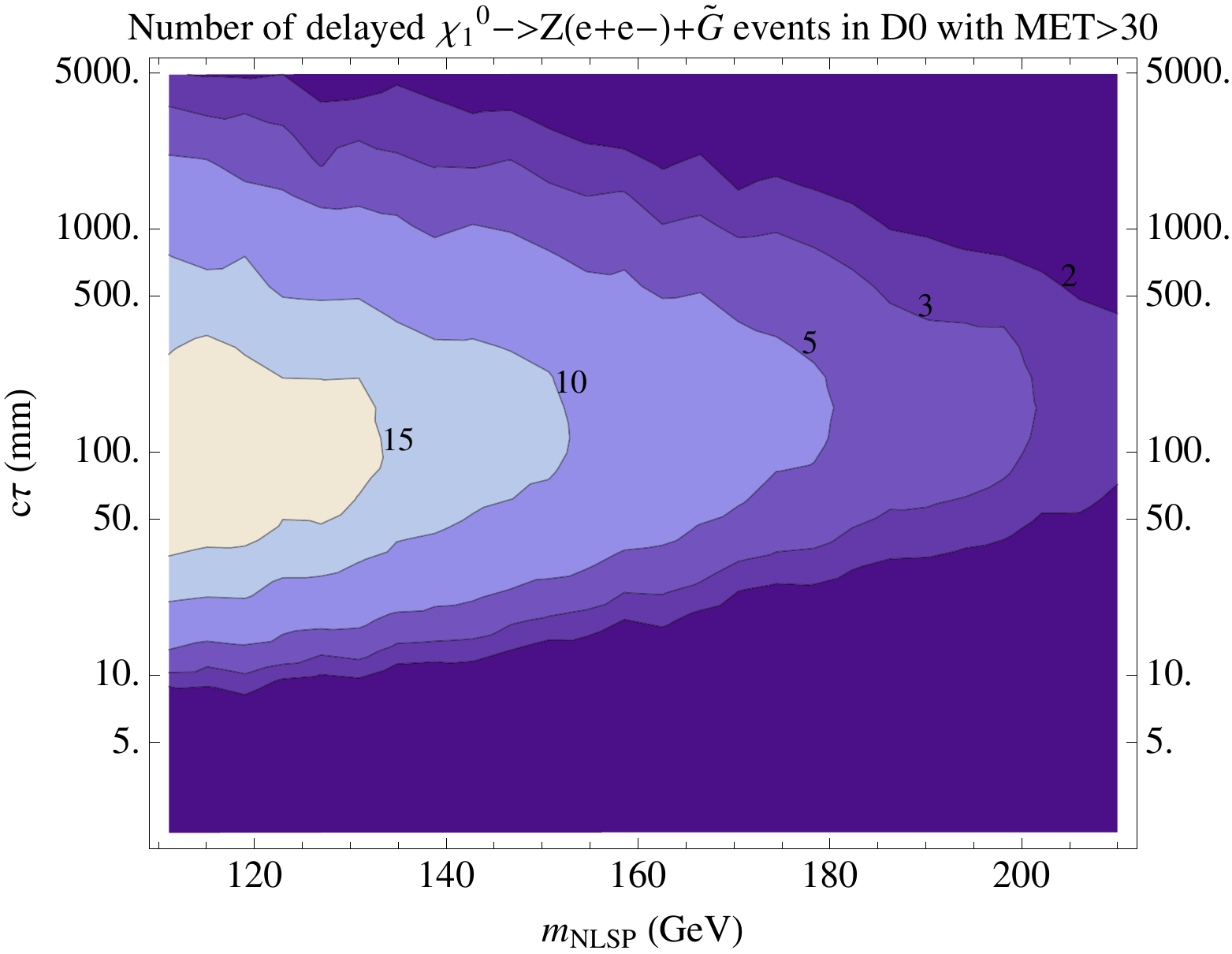}
\end{center}
\caption{Number of events at \dzero in 10 fb$^{-1}$, using the cuts from an existing study \cite{longzeeggd0} in combination with a MET cut of 30 GeV. }
\label{fig:d0reach}
\end{figure}

To illustrate this we first reproduced the acceptances given in~\cite{longzeeggd0} for $b'\rightarrow bZ(ee)$ using a Pythia simulation \cite{Pythia}.  We then calculated the number of events coming from direct production of a $Z$-rich Higgsino NLSP using the same cuts as in the previous study but including a MET cut of 30 GeV.\footnote{Note that in~\cite{longzeeggd0} the same MET cut was applied, but the invariant mass was only required to be above 20 GeV, which left 7 background events in 1.1 fb$^{-1}$.  For Higgsinos, the electrons will satisfy {\it both} a tight $Z$ mass window cut and a hard MET cut. We expect the combination of these will completely eliminate the background.}  In Figure~\ref{fig:d0reach} we plot the number of events in 10 fb$^{-1}$ at \dzero as a function of lifetime and mass.  We clearly see that \dzero has the potential to discover GMSB for Higgsino NLSPs for larger lifetimes, assuming the background can be made negligible. Note that since we are assuming only direct production of neutralinos and charginos, our result should be viewed as a conservative estimate of the reach, since any additional particles (sleptons, gluinos, squarks) will only add to the total rate for this inclusive search.

Although our paper is focused on signatures of displaced $Z\to \ell^+\ell^-$, for completeness' sake, let us also briefly discuss various Tevatron searches for other displaced final states.
 Recently \dz, motivated by Hidden Valleys~\cite{hv1,hv2}, has looked for pairs of neutral long lived particles that decay into $b\bar{b}$ pairs, both with displaced vertices~\cite{longlivebbd0}.  This study is sensitive to displacements between $1.6$ cm and $\sim 20$ cm. This study could be potentially very useful for bounding Higgs-rich Higgsino NLSPs.  Unfortunately, QCD is the background for this search and it is quite large.  Once again, however, this search did not include a MET cut. Including a hard MET cut could reduce the background significantly and with more data could turn this into a discovery mode.   

Using an EM calorimeter timing system with a resolution of about 1 ns \cite{cdfemtiming,TobackWagner}, CDF has searched for long lived neutralinos in GMSB that decay into photons at large displacements in~\cite{cdfgammajetmettiming}. In minimal gauge mediation, this has its best reach for lifetimes of about 150 cm, excluding a bino mass of about 105 GeV. This search could conceivably also be used to bound light Higgsino or wino NLSPs, where decays to heavier gauge bosons and Higgses are phase-space suppressed.  Clearly, further studies using the EM timing system of CDF would be useful and would be  complementary to \dz's studies. Interesting searches using EM timing that could probe long-lived general neutralino NLSPs include searches for out-of-time tracks or jets. Another possibility would be to search in the single photon plus $\met$ channel. Making this an exclusive final state (i.e.\ vetoing on other activity in the event) would be an interesting way to look for light Higgsino or wino NLSPs, where one decays outside the detector, while the other decays inside to $\gamma+\tilde G$. Since Higgsinos and winos (in contrast to binos) can be directly produced, there would not be complicated SUSY cascade decays giving rise to many high $p_T$ particles, and hence not much activity in the event other than the single photon plus missing energy.

\end{document}